\documentclass[11pt]{article}
\pdfoutput=1
\usepackage{amsmath}
\usepackage{amssymb}
\usepackage{graphicx,bbm,mathrsfs}
\usepackage{nicefrac}
\usepackage{bbm}
\usepackage{geometry}       
\usepackage{jheppub}     

\usepackage{dcolumn}   % needed for some tables
\usepackage{bm}        % for math
\usepackage{graphicx,mathrsfs}
\usepackage{nicefrac}
\usepackage{multirow}
\usepackage{color}

\def\ie{{\it i.e.}}

\newcommand{\be}{\begin{equation}}  
\newcommand{\ee}{\end{equation}}  
\newcommand{\bea}{\begin{eqnarray}}  
\newcommand{\eea}{\end{eqnarray}}  

\def\gev{\, {\rm GeV}}

\newcommand{\tr}{\operatorname{tr}}

\DeclareRobustCommand{\fbi}{\ensuremath{\mathrm{fb}^{-1}}}
\DeclareRobustCommand{\pt}{\ensuremath{p_{\mathrm{T}}}}

\renewcommand{\Re}{\operatorname{Re}}
\renewcommand{\Im}{\operatorname{Im}}

\newcommand{\arcosh}{\operatorname{arcosh}}
\newcommand{\Li}{\operatorname{Li_2}}
\newcommand{\nn}{\nonumber}

\newcommand{\specialcell}[2][c]{%
  \begin{tabular}[#1]{@{}c@{}}#2\end{tabular}}

\begin{document}

\vspace*{1.2cm}

\begin{center}

\thispagestyle{empty}
{\Large\bf Light-by-light scattering with intact protons at the LHC: from Standard Model to New Physics}\\[10mm]

\renewcommand{\thefootnote}{\fnsymbol{footnote}}

{\large  Sylvain~Fichet$^{\,a}$, Gero von Gersdorff$^{\,b\,}$, 
Bruno Lenzi$^{\,c\,}$, Christophe Royon$^{\,d\,}$, \\ 
Matthias Saimpert$^{\,d\,}$
\footnote{sylvain.fichet@lpsc.in2p3.fr,  gersdorff@gmail.com,  bruno.lenzi@cern.ch,  christophe.royon@cea.fr, matthias.saimpert@cern.ch} }\\[10mm]

\addtocounter{footnote}{-1} 

{\it
$^{a}$~International Institute of Physics, UFRN, 
Av. Odilon Gomes de Lima, 1722 - Natal-RN, Brazil \\
}
{\it $^b$ ICTP South American Institute for Fundamental Research, Instituto de Fisica Teorica,\\
Sao Paulo State University, Brazil \\
}
{ \it $^c$ CERN, CH-1211 Geneva 23, Switzerland \\
 }
{ \it $^d$  IRFU/Service de Physique des Particules, CEA/Saclay,\\ 91191 Gif-sur-Yvette cedex, France \\
 }

\vspace*{12mm}

{  \bf  Abstract }
\end{center}

We discuss the discovery potential of light-by-light scattering at the Large 
Hadron Collider (LHC), induced by the Standard Model (SM) and by  
new exotic charged particles. 
Our simulation relies on intact proton detection in the planned forward  detectors of CMS and 
ATLAS. 
The full four-photon amplitudes generated by any electrically charged 
particles of spins $1/2$ and $1$, including the SM processes involving loops of leptons, quarks and $W$ bosons are implemented in the Forward Physics Monte Carlo 
generator. 
Our method provides model-independent bounds on massive charged particles, only parametrized by the spin, mass and  
``effective charge'' $Q_{\rm eff}$ of the new particle.
We find that a new charged vector (fermion) with  $Q_{\rm eff}=4$ 
can be discovered up to $m=640$~GeV ($m=300$~GeV)  with an integrated luminosity of $300$ fb$^{-1}$ at the LHC. 
We also discuss the sensitivities to neutral particles such as a strongly-interacting heavy dilaton and 
 warped Kaluza-Klein gravitons, whose effects could be discovered for masses in the multi-TeV range.

\noindent
\clearpage

\section{Introduction}\label{se:intro}

A strong theoretical prejudice exists that New Physics (NP) beyond the 
Standard Model (SM) should appear around the TeV scale. 
However, after the first LHC run, a certain amount of popular models has 
been ruled out or they are cornered in fine-tuned regions of their parameter 
space. While the next LHC run is coming, it is more than ever important to be 
prepared to search for any kind of NP in the most possible robust ways.

A lot of models of physics beyond the SM predict the existence of new heavy particles with exotic electric charges. 
This happens for example in composite Higgs models, which require the existence of new charged particles of spin $\frac{1}{2}$ and $1$. In particular, 
the large mass of the top quark requires the existence of a composite top partner mixing with the elementary one.  
As the composite sector typically possesses a large global symmetry group, these top partners are accompanied by other resonances with exotic electric charges such as $\frac{5}{3}$ and $\frac{8}{3}$ \cite{Agashe:2004rs}.
New  particles with exotic electric charges can also appear 
in warped extra-dimension models with custodial symmetry \cite{Agashe:2003zs}.

Certain types of these particles are already constrained by direct searches 
at the LHC. 
Such direct searches are powerful in specific cases, but are highly 
model-dependent. Indeed, the production cross sections, decay chains and 
branching ratios all depend in general on  the details of the model. 
Therefore a specific analysis has to be tailored in each case, as both the 
final states and the backgrounds are  specific to the chosen model.  
Production cross sections at the LHC also vary greatly, depending crucially 
on whether or not the new states carry color. Placing general bounds 
on electrically charged particles in this way is thus not an easy task. 
Moreover, for certain  particles, the background can be large such that 
this type of search is not necessarily the most efficient one either.

\begin{figure}
\begin{center}
\includegraphics[scale=.75]{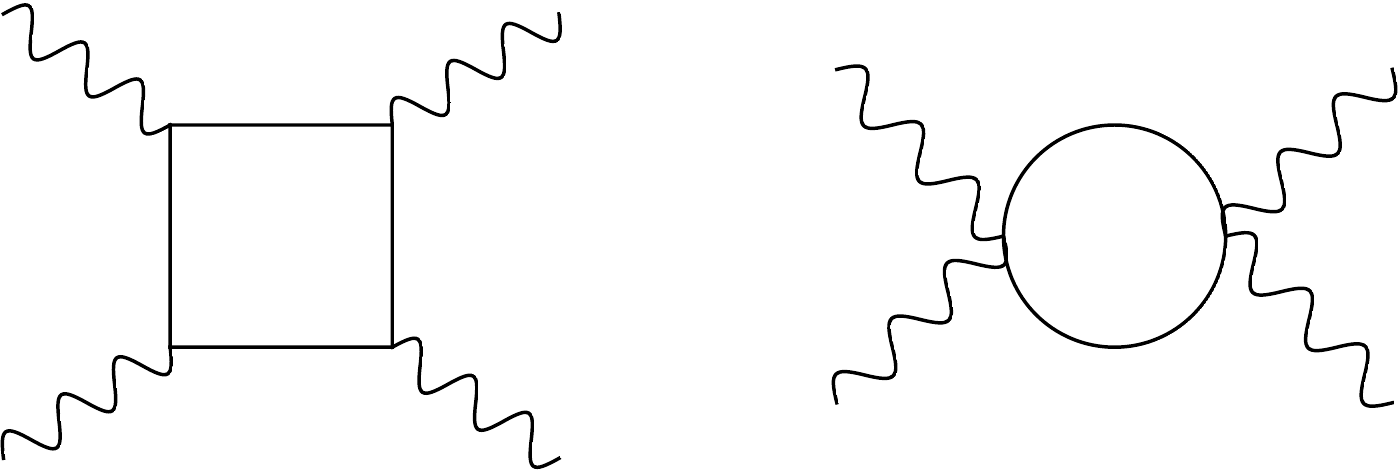}
\end{center}
\caption{\label{fig:loop}
Typical diagrams of electrically charged particles contributing to light-by-light scattering.
}
\end{figure}

In this paper we rather follow an alternative route, the one of  precision physics. 
We use the fact that any of such new electrically charged particles contribute to the scattering of light-by-light. This happens through a loop  as shown in Fig.~\ref{fig:loop}. 
Contrary to LHC direct searches, light-by-light scattering amplitudes are fully characterized by the mass, spin and electric charge of the  particle in the loop \cite{Fichet:2013ola,Fichet:2013gsa}. 
This property offers a way to search for charged NP in a fully  model-independent way.

In previous works \cite{Fichet:2013ola,Fichet:2013gsa}, such contributions were only included in an effective Lagrangian framework where they were matched to local effective operators of the type $F_{\mu\nu}^4$. 
The obvious drawback of this approach is that either one has to consider particles of masses much larger than the typical di-photon energy at the LHC -- in which case the sensitivity is poor -- or one has to introduce ad-hoc form-factors to mimic the unknown amplitude behaviour near the threshold. 
 In this work we go beyond this effective operator approach, and consider the full one-loop amplitudes to light-by-light scattering from NP. This allows us to obtain reliable estimates for the LHC sensitivity for NP particles of any mass. In the high mass limit the results coincide with those obtained previously \cite{Fichet:2013gsa}.  
The one-loop amplitudes are implemented in the Forward Physics Monte Carlo (FPMC)  generator~\cite{FPMC} that we use in the simulation.

 The SM quarks and leptons as well as the $W$ boson also contribute to the 
 light-by-light scattering via loops. One should notice that, at the LHC, typical di-photon energies are much larger than the masses of these particles. However, while the fermion loop amplitudes approach constants at high energies, the $W$ loop amplitude grows logarithmically \cite{Jikia:1993tc}. It is therefore the main SM contribution to light-by-light scattering at the LHC whereas it is rarely included in the background simulations. 

It is fairly surprising that prospects for studying light-by-light scattering  
are good at a hadron collider. This potential relies on the forward 
proton detectors that are planned to be built at about 220 m from the ATLAS main
detector within the AFP 
project~\cite{atlas}. The CMS and TOTEM collaborations
plan to use their forward proton detectors located at about the same position (CT-PPS project)\cite{cms}.
Using these detectors, four-photon interactions can be detected  with an 
unprecedented precision. Previous studies using proton-tagging at the LHC 
for New Physics searches can be found in 
Refs.~\cite{usww, usw,Sahin:2009gq,Atag:2010bh, Gupta:2011be, Lebiedowicz:2013fta,
Fichet:2013ola,Fichet:2013gsa,Sun:2014qoa,
Sun:2014qba,Sun:2014ppa,Sahin:2014dua,Inan:2014mua}.
We refer to \cite{friend:2013yra} for a study of light-by-light scattering at the LHC without proton tagging.

The outline of this paper is as follows. We first describe the exclusive di-photon production predicted by the Standard Model and discuss potential SM measurements at the LHC in Sec.~\ref{se:sm}. In Sec.~\ref{se:theory} we detail new charged particles contributions to the four-photon amplitudes both in the simple decoupling limit as well as the full energy range. Sections~\ref{se:fwd} and \ref{se:simu} are dedicated to the forward proton detectors and the implementation of the simulation. Backgrounds and cuts for the simulation are detailed in Sec~\ref{se:bg} and the expected sensitivity to new charged particles at the 14 TeV LHC is given is Sec.~\ref{se:lhc}. Finally,
the reach on other new physics candidates inducing light-by-light scattering is discussed in  Sec.~\ref{se:otherNP}.

\section{Standard Model exclusive di-photon production}\label{se:sm}

\subsection{Equivalent photon approximation}

We use the  the Equivalent Photon Approximation 
(EPA)~\cite{Terazawa:1973tb,Budnev:1974de} to describe the two-photon production in $pp$ 
collision. The almost real photons (with low virtuality 
$Q^2=-q^2$) are 
emitted by the incoming protons producing an object $X$, $pp\rightarrow pXp$, 
through two-photon exchange $\gamma\gamma\rightarrow X$.
The photon spectrum of virtuality $Q^2$ and
energy $E_{\gamma}$ is proportional to the Sommerfeld fine-structure 
constant $\alpha$  and reads:
\begin{equation}
d N = \frac{\alpha_{em}}{\pi}\frac{d E_{\gamma}}{E_{\gamma}}\frac{d Q^2}{Q^2}
 	 \left[ \left(1-\frac{E_{\gamma}}{E}\right)\left(1-\frac{Q^2_{min}}{Q^2}\right)F_E +
	         \frac{E_{\gamma}^2}{2E^2}F_M\right]
\label{eq:flux}
\end{equation}
where $E$ is the energy of the incoming proton of mass $m_p$, $Q^2_{min}\equiv
m^2_p E^2_{\gamma}/[E(E-E_{\gamma})]$ the photon minimum virtuality allowed by
kinematics and $F_E$ and $F_M$ are functions of the electric and magnetic form factors. They read
in the dipole approximation~\cite{Budnev:1974de,usww} 
\begin{equation}
F_M=G^2_M  \qquad F_E=(4m_p^2G^2_E+Q^2G^2_M)/(4m_p^2+Q^2)\qquad G^2_E=G^2_M/\mu_p^2=(1+Q^2/Q^2_0)^{-4}
\label{eq:newera:elmagform}
\end{equation}
The magnetic moment of the proton is $\mu_p^2=7.78$ and the fitted scale $Q^2_0=0.71$ GeV$^2$.
Since the electromagnetic form factors fall steeply as a function of $Q^2$,
the two-photon cross section can be factorized into the sub-matrix element and 
the two photon 
fluxes. In order to obtain the production cross section, the photon fluxes are first integrated over $Q^2$
\begin{equation}
f(E_\gamma)=\int^{Q^2_{max}}_{Q^2_{min}}\frac{d N}{d E_\gamma d Q^2} d Q^2
\label{sm:flux_q2}
\end{equation}
up to a sufficiently 
large value of $Q^2_{max}\thickapprox2-4$ GeV$^2$. The result 
is given for instance in Ref.~\cite{usww}.

The contribution to the integral above $Q^2_{max}\thickapprox$ 2 GeV$^2$  is very small. 
The $Q^2$-integrated photon flux also falls rapidly as a function of the photon energy $E_{\gamma}$ 
which implies that the two-photon production is dominant at small masses 
$W\approx2\sqrt{E_{\gamma1}E_{\gamma2}}$. 
Integrating the product of the
photon fluxes $f(E_{\gamma1})\cdot f(E_{\gamma2})\cdot d E_{\gamma1}\cdot 
d E_{\gamma2}$ from both protons over the photon  energies while keeping
the two-photon invariant mass fixed to $W$, one obtains the two-photon effective
luminosity spectrum $d L^{\gamma\gamma}/d W$.

The production rate of massive objects via photon exchange at the LHC
is however limited by the photon 
luminosity at high invariant mass. The integrated two-photon 
luminosity above $W>W_0$ for $W_0=23$ GeV,\ $2\times m_W\thickapprox160$ GeV, 
and 1 TeV is respectively $1\%$, $0.15\%$ and $0.007\%$ of the luminosity 
integrated over the whole mass spectrum.

Using the effective relative photon
luminosity $d L^{\gamma\gamma} \slash d W$, the total cross section reads 
\begin{equation}
\sigma=\int\sigma_{\gamma\gamma\rightarrow X}
      \frac{d L^{\gamma\gamma}}{d W}d W 
\label{eq:sm:totcross}
\end{equation}
where $\sigma_{\gamma\gamma\rightarrow X}$ denotes the 
cross section of the sub-process $\gamma\gamma\rightarrow X$, dependent on the 
invariant mass of the two-photon system.

In these studies, we assume both protons to be intact after interaction.
Additional soft gluon exchanges between the two protons might destroy the
protons. A traditional way to take this effect into account is to introduce the
so-called survival probability that the protons remain
intact~\cite{survival1,survival2}
in $\gamma$ induced processes.
In our studies, we assumed a survival probability of about 90\%. More recent
studies~\cite{Dyndal:2014yea} show that this might be slightly optimistic and the
survival probability at high di-photon masses might be of the order of 60\%. In
that case, the yield should be reduced accordingly. It is thus important 
to measure that quantity at the LHC.

\subsection{Standard Model exclusive di-photon production}

\begin{figure}
\centering
\includegraphics[scale=0.3]{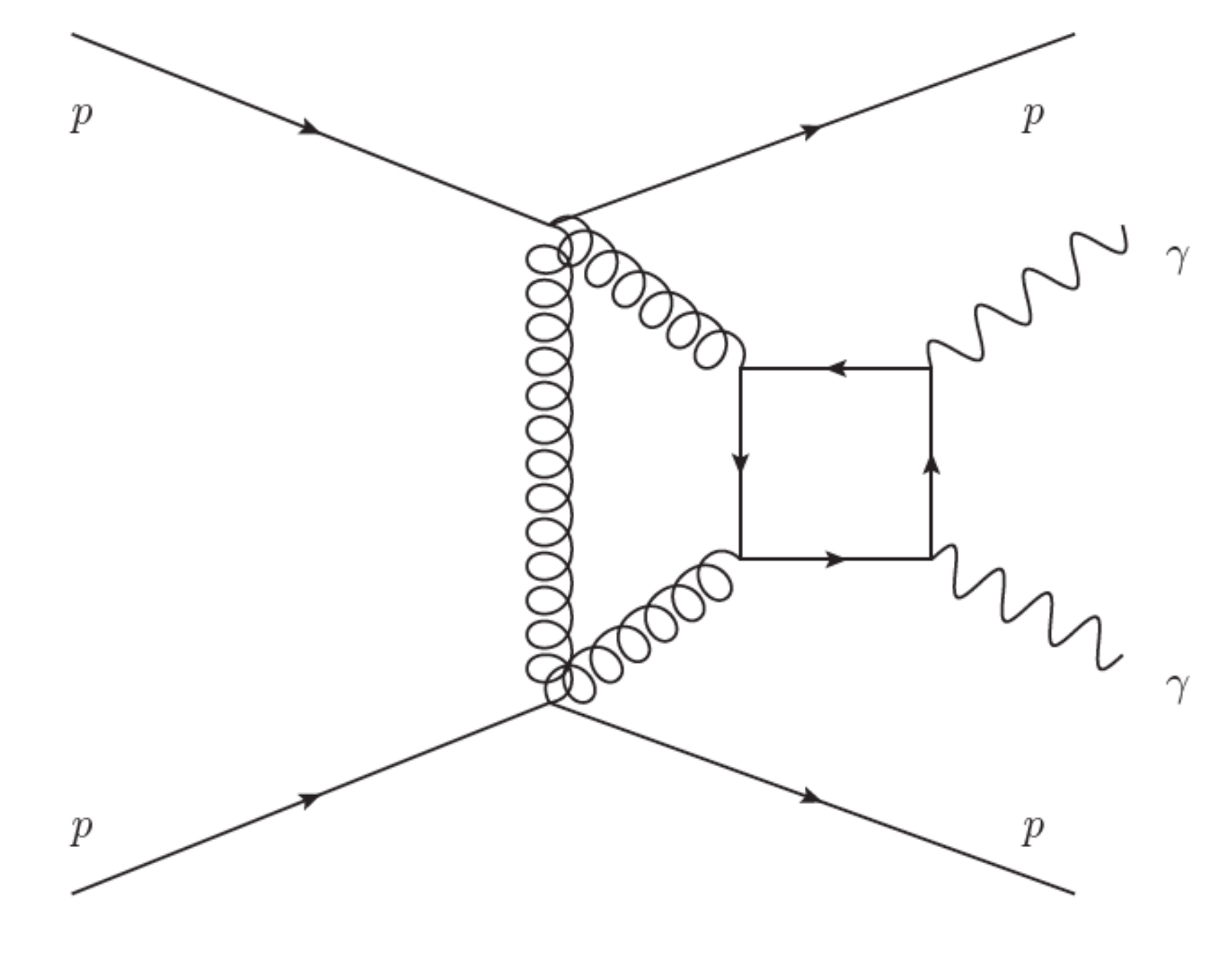}
\includegraphics[scale=0.3]{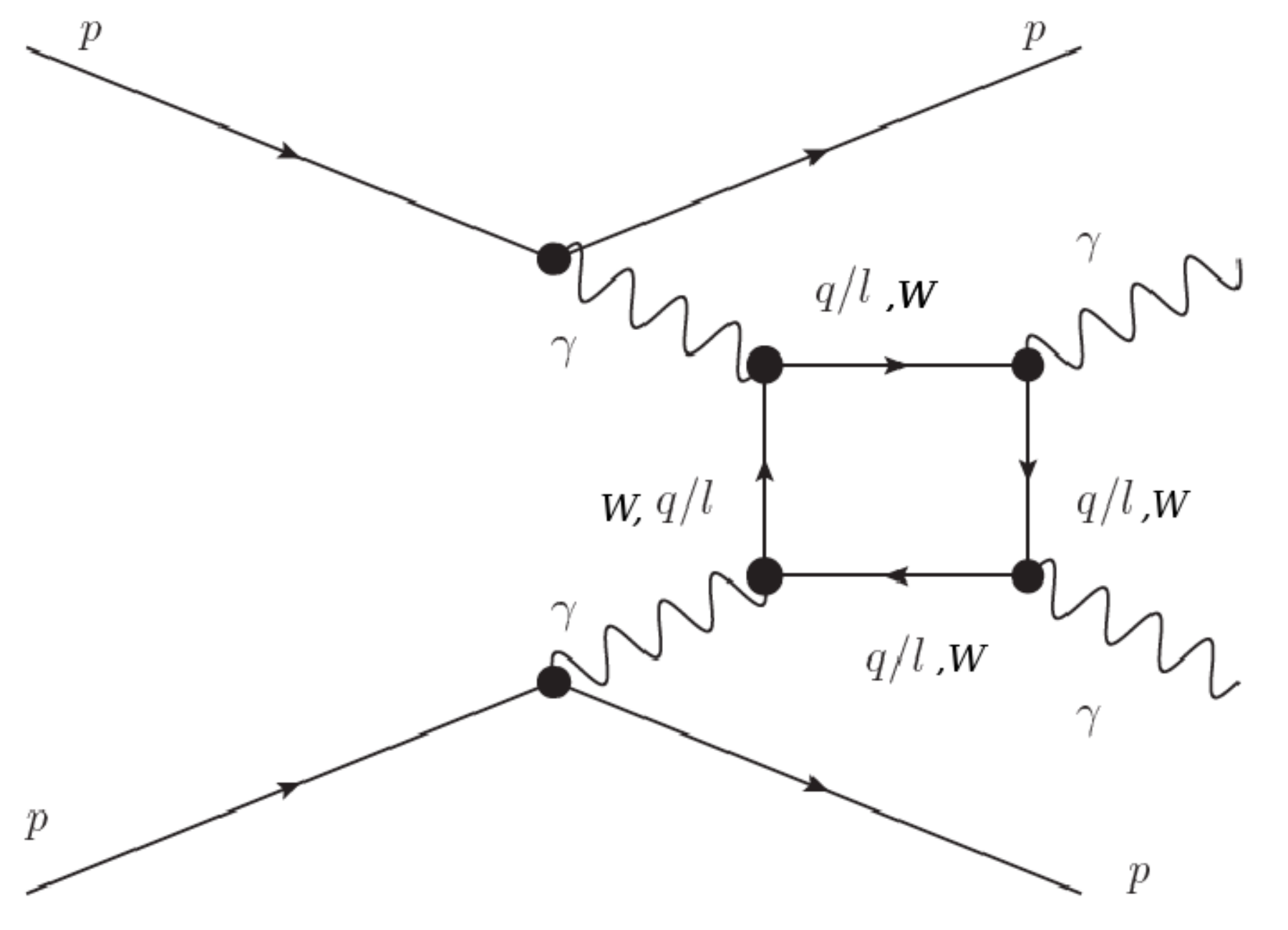}
\caption{Feynman diagrams predicted by the Standard Model leading to the exclusive production of two photons and two intact protons in the final state at the lowest order of perturbation theory. }
\label{diag}
\end{figure}

The Standard Model predicts exclusive di-photon production with two intact protons through various 
processes, which can be decomposed in two parts, as shown in Fig.~\ref{diag}. 
The first diagram (Fig.~\ref{diag}, left) corresponds to exclusive QCD 
di-photon
production via gluon exchanges~\cite{Khoze:2001xm}  (the second gluon ensures that 
the exchange is
colorless leading to intact protons in the final state) and the second one 
(Fig.~\ref{diag}, right) via
photon exchanges.

\begin{figure}
\begin{center}
\includegraphics[scale=0.5]{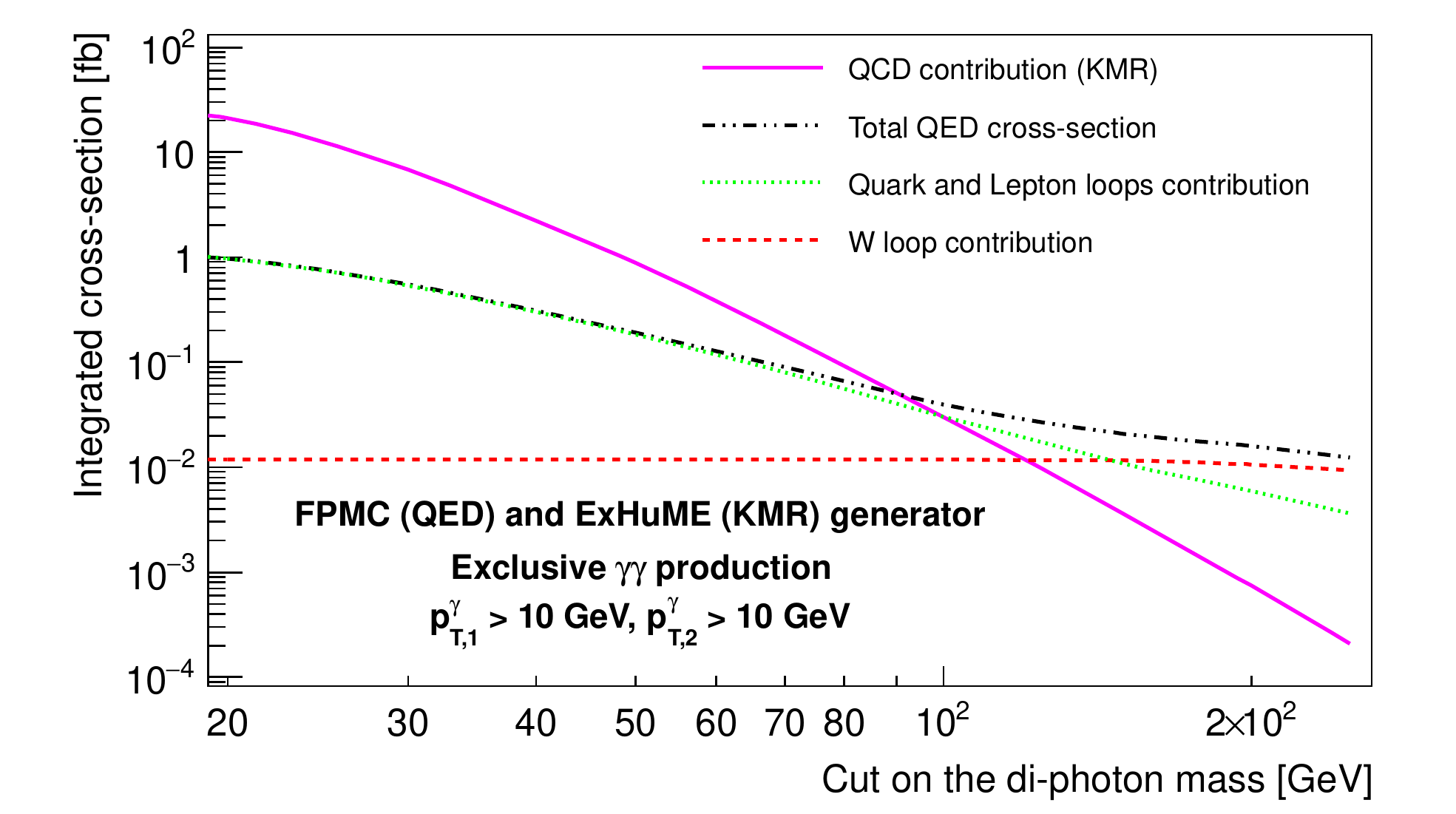}
\end{center}
\caption{Integrated cross sections of the different exclusive di-photon processes with intact protons at the $13$ TeV LHC, plotted against the required minimum di-photon mass. Both photons are required to have a transverse momentum above 10 GeV.}
\label{matthias}
\end{figure}

It is worth noticing that quarks, leptons and $W$-boson loops plus the associated interference terms need to be considered in order to get the correct SM cross section. 
These loops have been computed in Refs.~\cite{Karplus:1950zza,Karplus:1950zz,Costantini:1971cj,Jikia:1993tc}. We collect explicit expressions in App.~\ref{app:Amplitudes}.
The various contributions are  illustrated in Fig.~\ref{matthias}, where we display the integrated cross sections
of the different exclusive di-photon processes varying the requirement on the minimum mass of the photon pair. Both photons are required to have a transverse momenta above 10 GeV. 
We can see that the QCD induced processes
are dominant at low di-photon mass whereas
the photon induced ones (QED processes) dominate at higher masses. The quark and lepton loops contribution is the second leading one,
whereas the $W$ loop contribution dominates at very high di-photon 
masses. The QED contribution starts getting over the QCD one as of a di-photon mass
of $\sim$100 GeV.  The $W$-loop contribution is omitted in most of the
studies and it is one of the first times that all terms (including interference) are implemented in a single MC generator,
FPMC.

In Table~\ref{tab_smprod}, we report a few values of the cross sections from Fig.~\ref{matthias} discussed above. The threshold where the QED contribution starts
dominating the production is for a di-photon mass slightly below 100 GeV as we
already mentioned. A low mass measurement of the gluon contribution
would be possible at the LHC in the case of the special runs at low luminosity and low  pile up ($\mu\sim 1-2$)
with modified optics ($\beta^*=90$ m), currently being discussed among the 
different LHC experiments \cite{LHCC}. The intact protons would be detected in the vertical roman  pots of the TOTEM or ALFA
detectors. Thanks to the low instantaneous luminosity of those special runs 
one should be able to implement a dedicated di-photon trigger with 
$p_{T}$ thresholds as low as 
$p_{T1,2}>$ 5 GeV, and following Table~\ref{tab_smprod}, a di-photon mass requirement of $m_{\gamma \gamma}>$10 GeV would lead to a sizeable
cross section of about 370 fb. For  a typical integrated luminosity of 
0.1 fb$^{-1}$ expected in these
special runs, which corresponds approximately to a week of data taking, 37 events can be measured and compared to the different exclusive di-photon
cross section calculation  and to the previous and unique 
measurement of this process done by the CDF collaboration~\cite{cdf}.

On the other hand, the Standard Model QED production does not seem to be
reachable at the LHC in $pp$ collisions. It might be possible to
study the di-photon production via quark, lepton and even $W$ loops at the LHC in
the heavy ion mode~\cite{friend:2013yra,eloi}.

\begin{table*}
\begin{center}
\small
\begin{tabular}{|c|c|c|c|}
\hline
Cut / Process & QCD Exclusive (KMR) & QED Fermion loop & $W$ loop \\
\hline
\hline
$m_{\gamma\gamma}>10$~GeV,$p_{T1,2}>5$~GeV       & 372.1  fb    & 5.5   fb   & 0.012  fb  \\
\hline
\hline
$m_{\gamma\gamma}>20$~GeV,$p_{T1,2}>10$~GeV      & 20.4   fb    & 1.0   fb   & 0.012  fb  \\
$m_{\gamma\gamma}>50$~GeV,$p_{T1,2}>25$~GeV      & 0.87   fb    & 0.18  fb   & 0.012  fb  \\
$m_{\gamma\gamma}>100$~GeV,$p_{T1,2}>50$~GeV     & 0.030  fb    & 0.03  fb   & 0.012  fb  \\
$m_{\gamma\gamma}>200$~GeV,$p_{T1,2}>100$~GeV     & 7.4e-4 fb    & 5.0$\cdot$10$^{-3}$ fb   & 0.010 fb  \\
$m_{\gamma\gamma}>500$~GeV,$p_{T1,2}>250$~GeV     & 3.2e-6 fb    & 3.0$\cdot$10$^{-4}$ fb   & 0.004 fb  \\
\hline
\end{tabular}
\end{center}
\caption{Integrated cross sections of the different SM exclusive di-photon production processes at the LHC at $\sqrt{s} = 14$ TeV for various requirements on the di-photon mass (m$_{\gamma\gamma}$) and photon transverse momenta (p$_{T1,2}$).}
\label{tab_smprod}
\end{table*}

\section{Effects of new charged particles on exclusive di-photon production}\label{se:theory}

\subsection{General Considerations}

\label{se:general}

The particles running in the loops in Fig.~\ref{fig:loop} are characterized by their electric charge $Q$, their spin $S$, and their mass $m$. The loop amplitude is  proportional to $\alpha^2_{\rm em}Q^4$.

However, the new particles also have  in general a 
multiplicity with respect to electromagnetism. For instance, the multiplicity is
three if the particles are colored.
%If the mass splitting due to electoweak breaking is negligible, 
One can simply take into account this multiplicity by defining
\be
Q_{\rm eff}^4=\tr Q^4
\ee
where the trace goes over all particles with the same approximate mass.
The amplitude then becomes proportional to 
\be \mathcal{M}\,\propto \alpha^2_{\rm em} Q_{\rm eff}^4\,.\ee

As an example, consider the minimal composite Higgs models with global 
symmetry group $G=SO(5)\times U(1)_X$ \cite{Agashe:2004rs}. The simplest and 
most common choices for the embeddings of the quark partners are 
the  $5_{\frac{2}{3}}$ or $14_\frac{2}{3}$ representations of $G$.
After the breaking $G\to H= SU(2)_L\times SU(2)_R\times U(1)_X$ the theory predicts light vector-like (VL) fermions in the $(1,1)_\frac{2}{3}$,  $(2,2)_{\frac{2}{3}}$ or $(3,3)_\frac{2}{3}$ representations. The latter two actually contain states of various electric charges which are approximately degenerate in mass. One then obtains \footnote{The effective charges are dominated by the states of highest charge in each case, which are a single $Q=\frac{5}{3}$ ($Q=\frac{8}{3}$) state in the case of the $(2,2)$ and $(3,3)$ respectively.}
\bea
Q_{\rm eff}&=&2.22\qquad (2,2){\rm\ representation}\nn\\
Q_{\rm eff}&=&3.80\qquad (3,3){\rm\ representation}
\eea
which in particular  contain the multiplicity due to color.
For larger global groups the effective charges can become even larger, as several of the above mentioned representations can occur simultaneously. 
There is in principle no reason to restrict to the smallest group $SO(5)$ and to the smallest representations mentioned above, other than simplicity and minimality. Larger representations raise the effective charge more efficiently than going  to larger groups.

Various bounds on  VL quarks from direct searches already exist. VL quarks mixing strongly with third generation quarks (``top/bottom partners") are most strongly constrained and yield bounds of the order of $400-700$ GeV, depending on the branching ratios.   
Bounds are generally weaker for VL quarks mixing with lighter generations \cite{Barducci:2014ila}. 
To the best of our knowledge there are not yet any bounds on VL leptons, which are equally predicted in many of these models. Note that VL leptons have much smaller production cross sections at the LHC.

Summarizing, while direct searches already constrain quite a lot the parameter space of specific models with VL quarks, it is difficult to extract completely model independent bounds, and some assumptions about specific couplings/mixings or branching fractions are required. On the other hand, our results will be expressed in terms of mass, spin and effective charge $ Q_{\rm eff}$, and any  specific model can be easily mapped onto these quantities.

\subsection{Effective Field Theory (EFT)}
\label{sec:eft}

In the limit where the mass of the new charged particle is large with respect to the energy of the process, $m\gg E$,  one can describe the four-photon interactions using higher-dimensional local operators in an effective Lagrangian,  
\be
\mathcal{L}_{4\gamma}= %\frac{\zeta_1^\gamma}{\Lambda^4} 
\zeta_1 F_{\mu\nu}F^{\mu\nu}F_{\rho\sigma}F^{\rho\sigma}
+\zeta_2 F_{\mu\nu}F^{\nu\rho}F_{\rho\lambda}F^{\lambda\mu}
\label{zetas} \,.
\ee
These are operators of dimension 8. The coefficients $\zeta_1$, $\zeta_2$, although they can be studied separately as in the previous work \cite{Fichet:2013gsa}, are ultimately both predicted by any model of New Physics. 
From the effective Lagrangian Eq.~(\ref{zetas}) one can compute the 
unpolarized four-photon angular cross section  
\be
  \frac{d\sigma}{d\Omega}
  =\frac{1}{16 \pi^2\,s}(s^2+t^2+st)^2
  \left[48 (\zeta_1)^2 + 40 \zeta_1 \zeta_2 + 11 (\zeta_2)^2\right]
  \label{xsec}
  \ee
where $s$, $t$ are the usual Mandelstam variables. 
Since the amplitudes interfere with the background (e.g.~the $W$ loops), 
Eq.~(\ref{xsec}) is valid at sufficiently large $s$ or $\zeta_i$. 
Let us stress that Eq.~(\ref{xsec}) only relies on the effective Lagrangian Eq.~(\ref{zetas}) and makes no reference to the origin of the coefficients $\zeta_i$.

As the EFT is nonrenormalizable, at high energies one expects a breakdown of unitarity. Using the well-known partial wave analysis \cite{Agashe:2014kda} we can estimate for what values of $\zeta_i$ and $s$ the theory remains unitary. By imposing unitarity on the $S$-wave of the EFT  amplitudes in Eq.~(\ref{amplitudesEFTlimit}) one finds the conditions
\bea
(4\zeta_1+3\zeta_3)s^2<4\pi\,,\qquad (4\zeta_1+\zeta_2)s^2<\frac{12}{5}\pi\,,
\eea
for the $\mathcal M_{++++}$ and $\mathcal M_{++--}$ amplitudes respectively. As most of the recorded diphoton events have $\sqrt{s}$ below 1 TeV (see Fig.\ref{accept}), we expect the EFT to remain unitary for couplings up to 
\be
\zeta_i\lesssim (10^{-12}-10^{-11}) {\rm\ GeV}^{-4}\,.
\label{unitarity}
\ee
The sensitivities we will derive in Sec.~\ref{se:effective} are much better than these unitarity bounds. However, we remark that unless the underlying New Physics model is very strongly coupled, the condition $ m<E  $ provides a stronger constraint on the $\zeta_i$ and EFT typically breaks down before unitarity is violated. Condition (\ref{unitarity}) should thus be considered as an absolute model-independent upper bound above which we no longer can trust the EFT approximation.

We now return to the case of new electrically charged particles with arbitrary spin $S$. 
 Using the background field method as in the general computation of  \cite{Fichet:2013ola}, we obtain the following expression for the coefficients of the $4\gamma$ operators, 
\be 
\zeta_i=\frac{\alpha^2_{\rm em} Q_{\rm eff}^4}{m^{4}}\,c_{i,S}\,,
\label{zetaEFT}
\ee 
where
\be
c_{1,S}=
\begin{cases}
\frac{1}{288} & S=0 \\
-\frac{1}{36} & S=\frac{1}{2} \\
-\frac{5}{32} & S=1 \\
\end{cases}
\,,\quad
c_{2,S}=
\begin{cases}
\frac{1}{360} & S=0 \\
\frac{7}{90} & S=\frac{1}{2} \\
\frac{27}{40} & S=1  \\
\end{cases} \quad.
\label{EH}
\ee
  The contributions from the scalar are smaller by one order of magnitude  
  with respect to the fermion and vector.
It can easily be checked that in the case of fermions $\mathcal L_{4\gamma}$ reduces to the famous Euler-Heisenberg Lagrangian
\cite{Heisenberg:1935qt}.

One may observe that the magnitude of the contributions to light-by-light scattering grows fast with the spin -- see Eq. \eqref{EH}. This intriguing fact suggests that contributions from higher-spin particles might be particularly large.  Higher-spin states potentially emerge in many extensions of the Standard Model. This includes the composite states arising from a new strongly-interacting gauge sector, as well as low-energy strings (see \cite{Anchordoqui:2014wha} for a recent review). 
A naive estimate using the background field method suggests that higher-spin contributions would go as $\zeta_i\propto S^5$. However this result cannot be fully trusted, because higher-spin Lagrangians intrinsically contain higher-dimensional interactions that  lead to extra-divergences in the loops. The tools necessary for higher-spin phenomenology are not yet available, and are under development \cite{fgwip}.

\subsection{Exact Amplitudes}  
  
The effective field theory analysis has the advantage of being very simple. 
However it is only  
 valid as long as the center-of-mass energy is small with respect to the 
 threshold of pair-production of real particles, $s\ll 4m^2$. 
Since the maximum proton missing mass (corresponding to the di-photon invariant mass in our case) is of the order of $\sim 2$ TeV at the 14 TeV LHC, for particles lighter than $\sim 1$ TeV the  effective field theory computation needs to be corrected.
This can be done by using ad-hoc form factors, as often done in the literature. 

The more correct approach is to take into account the full momentum dependence of the four-photon amplitudes.
They have been computed in the case of fermions in Ref.~\cite{Costantini:1971cj} and for vector bosons in Ref.~\cite{Jikia:1993tc}. 
Next-to-leading order corrections from QED and QCD are found to be negligible in \cite{Bern:2001dg}.

Following the notation and normalization of Ref.~\cite{Costantini:1971cj}, the unpolarized cross section can be expressed in terms of the various helicity configurations as
\bea
\frac{d\sigma}{d\Omega}
&=&\frac{\alpha_{\rm em}^4Q_{\rm eff}^8}{2\pi^2 s} \left(|\mathcal M_{++++}|^2+|\mathcal M_{++--}|^2+|\mathcal M_{+-+-}|^2\right.\nn\\
&&\left.+|\mathcal M_{+--+}|^2+4|\mathcal M_{+++-}|^2\right)
\eea
Due to the relations
\bea
\mathcal M_{+-+-}(s,t,u)&=&\mathcal M_{++++}(u,t,s)\,,\nn\\
\mathcal M_{+--+}(s,t,u)&=&\mathcal M_{++++}(t,s,u)
\label{relation}
\eea
only the configurations $\mathcal M_{++++}$, $\mathcal M_{++--}$ and $\mathcal M_{+++-}$ have to be computed.
We summarize them in App.~\ref{app:Amplitudes} together with various kinematical limits that are  useful both for theoretical as well as numerical considerations.
For comparison, we give in App.~\ref{app:C} the same amplitudes as obtained from the effective Lagrangian Eq.~(\ref{zetas}). All those amplitudes were implemented into a single MC generator dedicated to forward physics, the Forward Physics Monte Carlo Generator.

\section{The forward proton detectors} 

\begin{figure}
\begin{center}
\includegraphics[width=0.90\linewidth]{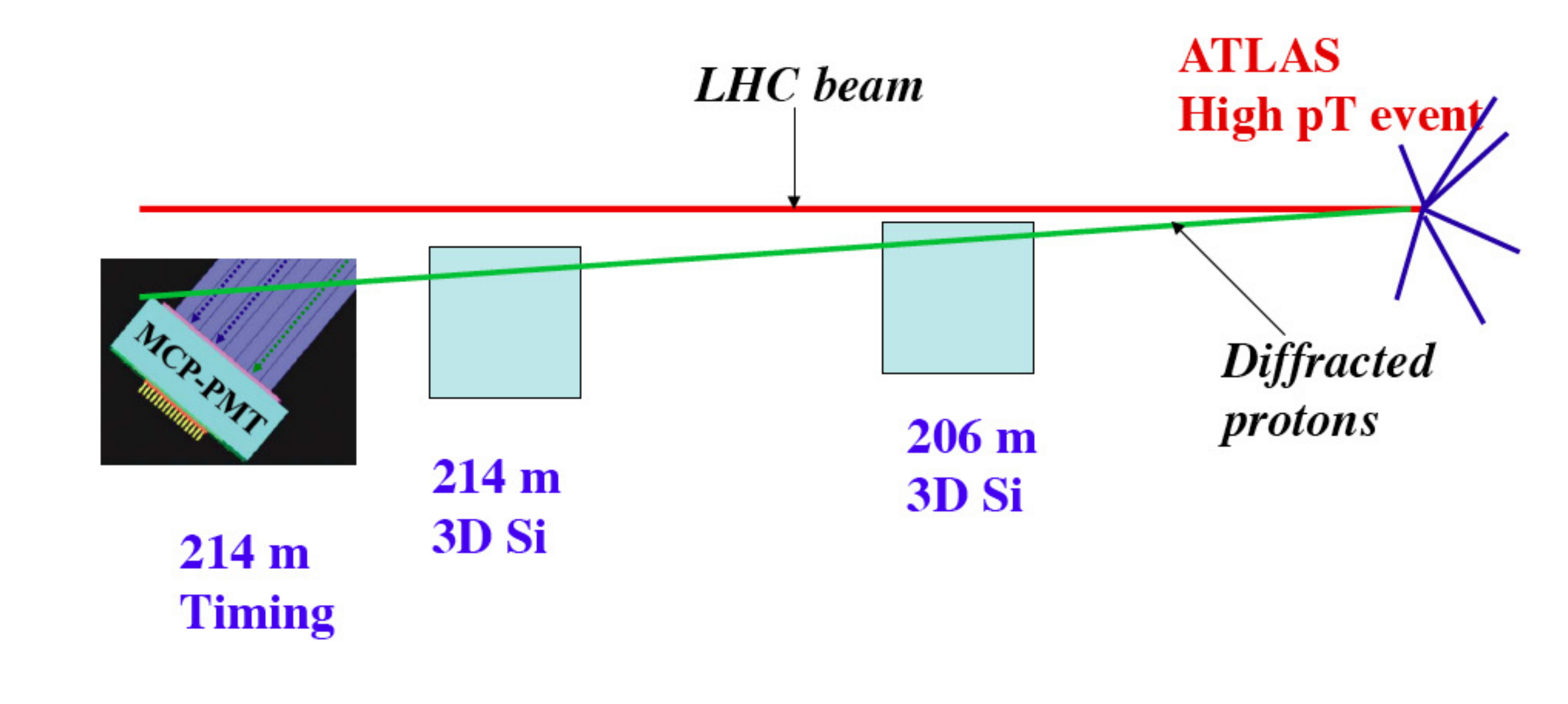}
\end{center}
\caption{Scheme of the AFP detector. Roman pot hosting Si and timing detectors
will be installed on both sides of ATLAS at 206 and 214 m from the ATLAS nominal interaction point. The
CMS-TOTEM collaborations will have similar detectors.}
\label{afp}
\end{figure}

In this study, the protons are assumed to be detected in 
the ATLAS Forward Proton (AFP) detector at 206 (AFP1 detector) and 214 (AFP2 detector)
meters on both sides of the ATLAS experiment~\cite{atlas} (see Fig.~\ref{afp}) 
or in a similar detector proposed by the TOTEM and CMS collaborations, 
the so called CT-PPS detector, 
to be installed on both sides of the CMS detector.

In AFP1, a tracking station composed by 6 layers of 
Silicon detectors will be deployed. The second section, AFP2, will contain a 
second identical tracking station and a timing detector.
Likewise, 
the CT-PPS of CMS will also use the same combination of tracking and timing 
detectors, with the far station using specially designed cylindrical roman 
pots to house the timing detectors~\cite{nicolo}. 

\begin{figure}
\begin{center}
\includegraphics[width=0.49\linewidth]{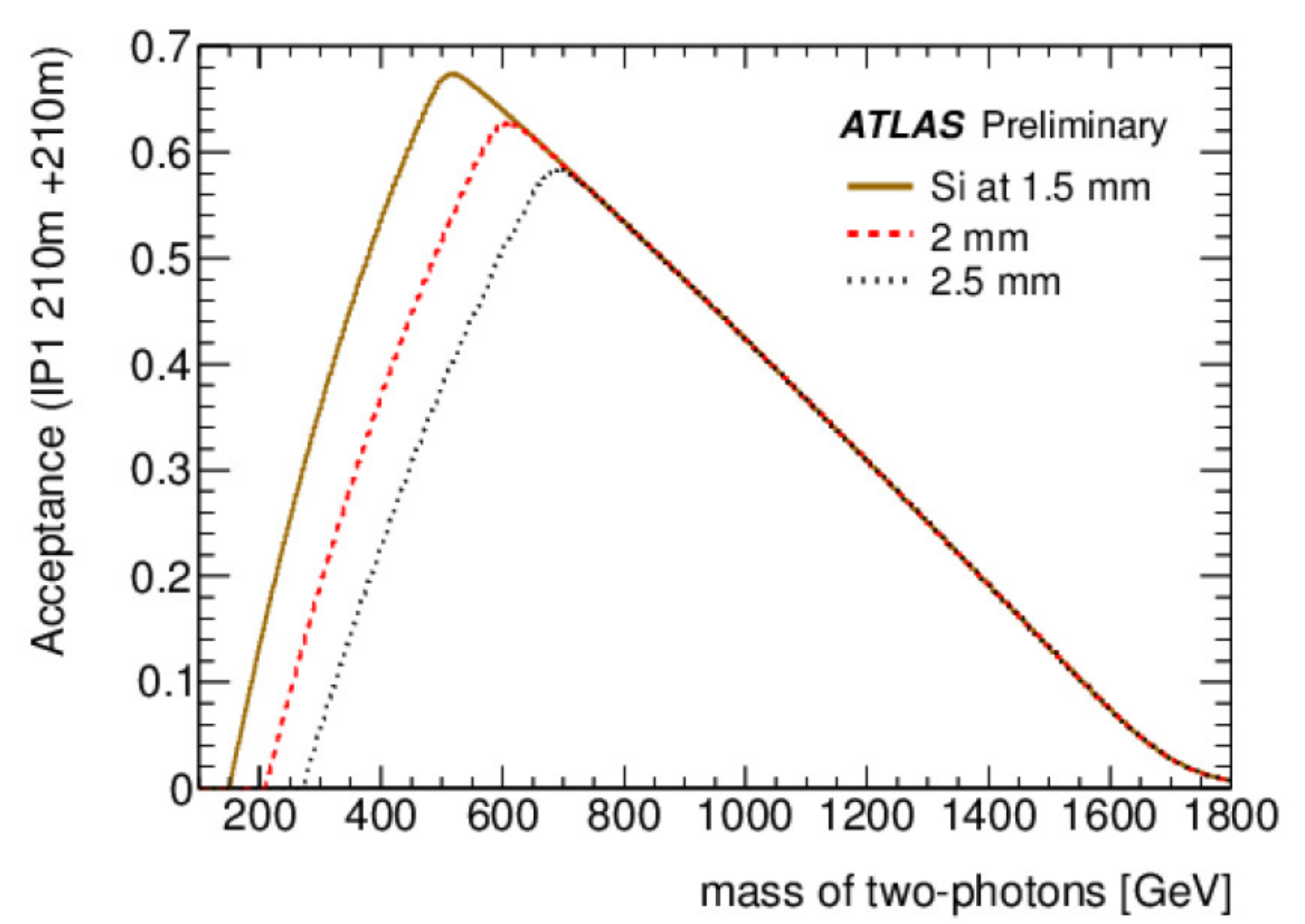}
\end{center}
\caption{Di-photon mass acceptance for the AFP detectors for the LHC 
nominal running at $\beta^*=$ 0.6 m.}
\label{accept}
\end{figure}

The proton taggers 
are expected to determine the fractional proton momentum loss $\xi$ in the 
range $0.015 < \xi < 0.15$ with a relative resolution of 2\%. This leads to the
acceptance in di-photon mass shown in Fig.~\ref{accept} between about 350 and
1700 GeV~\cite{atlas}.
In addition, 
the time-of-flight of the protons can be measured within 10~ps, which 
translates into 2.1~mm resolution on the determination of the interaction 
point along the beam axis $z$. In the following, we always assume both intact protons
in the final state to be tagged in AFP or CT-PPS.

\label{se:fwd}

\section{Setup of the simulation} 
\label{se:simu}

\subsection{Event generation with the Forward Physics Monte Carlo Generator} 

The Forward Physics Monte Carlo Generator was designed to produce single diffractive, double pomeron exchange (DPE), 
exclusive
diffractive and photon-induced processes within
the same framework. We use FPMC to produce all diffractive and photon
induced events, the diffractive and exclusive processes are implemented 
by modifying the \hbox{HERWIG}~\cite{herwig} routine for the 
 $e^+e^-\rightarrow(\gamma\gamma)\rightarrow X$ process. In case of the 
 two-photon $pp$ events, the Weizs\"{a}cker-Williams (WWA) formula describing 
 the photon emission off point-like electrons is substituted by the 
 Budnev flux \cite{Budnev:1974de} which describes properly the coupling of the photon to the proton, taking into account the proton electromagnetic structure. For the central exclusive production, a look-up table of the effective gluon-gluon luminosity computed by ExHuME \cite{ExHuME} is implemented. In case of the pomeron/reggeon exchange, the WWA photon fluxes are turned to the pomeron/reggeon fluxes multiplied by the diffractive parton density functions.

For processes in which the partonic structure of the pomeron is probed, 
the existing HERWIG matrix elements of non-diffractive production are used 
to calculate the production cross sections. The parton distributions in the
Pomeron as determined by the H1 collaboration at HERA (see \cite{Royon:2006by} an references therein) are used with a survival probability of 0.03~\cite{survival1,
survival2}.
The list of particles is corrected 
at the end of each event to change the type of particles from the initial 
state electrons to hadrons and from the exchanged photons to pomerons/reggeons, 
or gluons, depending on the process. 

\subsection{Background simulation and event reconstruction} 

In an experiment like ATLAS  or CMS, the photons can be reconstructed in the central detectors, instrumented with 
electromagnetic calorimeters covering the pseudorapidity range 
$|\eta| \lesssim 2.5$. The calorimeters provide excellent energy and 
position resolution, $\Delta E/E$ around 1\% for energies above few hundred 
GeV, $\Delta \eta \sim 0.001$ and $\Delta\phi \sim 1$~mrad to few mrad. For 
transverse momenta in the range of few tens of GeVs up to about 1 TeV 
and even in the presence of 100 additional collisions occurring in the same 
or neighbouring bunch crossings (pile up), the photon identification 
efficiency is expected to be around 75\% with jet rejection factors in 
excess of 4000 \cite{atlas_detector}. In addition, about 1\% of the electrons are mis-identified as
photons.  These numbers are for the ATLAS detectors but similar
sensitivities are expected for CMS.

A significant fraction of the photons convert to electron-positron pairs in 
the material upstream the calorimeters. In the region instrumented with 
silicon tracking detectors the material budget greatly varies as a function of 
$\eta$ in both experiments, typically between less than 0.5 radiation 
lengths ($X_0$) at $\eta = 0$ up to 2~$X_0$ at higher $\eta$. As a result, 
about 15-30\% of the photons convert in this region. The charged tracks associated 
contribute to the fake electron to photon rates 
that can reach 1\% and on the other hand can help locating the interaction 
point with sub-millimiter accuracy. An alternative method exploits the longitudinal 
segmentation of the ATLAS electromagnetic calorimeter to determine the photon 
production point along the beam axis within $\sim 15$~mm. 
By locating precisely the interaction point 
one can measure the photon trajectory and, in combination with the proton 
detectors, determine the four momenta of all particles in the final state. 
The constraint of the full event kinematics is an extremely powerful feature 
to reject the backgrounds where two photons are produced by a hard scattering 
process and two intact protons arise from  pile up interactions.

The analysis was designed to yield high signal selection efficiency and 
suppress the backgrounds that are divided into three classes. Exclusive 
processes with two intact photons and a pair of photon candidates include the 
SM light-by-light scattering, the central-exclusive production of two 
photons via two-gluon exchange and $\gamma\gamma \rightarrow e^+e^-$. 
Processes involving double pomeron exchange can result in protons 
accompanied by two jets, two photons and a Higgs boson that decay into two 
photons. Finally, one can have gluon or quark-initiated production of two 
photons, two jets or two electrons (Drell-Yan) with intact protons arising 
from  pile up interactions. Both the anomalous $\gamma\gamma \rightarrow 
\gamma\gamma$ signal, exclusive and DPE background processes were simulated by 
the FPMC generator, with the 
exception of the central exclusive production of $\gamma\gamma$ that was 
simulated using ExHuME.

\section{Event selection} 
\label{se:bg}

The number of expected signal and background events after each cut
is given in 
Table~\ref{tab:event} for an integrated luminosity of 300~\fbi\ ($\simeq$ 
3 years of data-taking at the LHC run 2) and 
50  pile up interactions for a center-of-mass energy of 14\,TeV.
We fix $S=1$, $Q_{\rm eff}=4$, $m=340$ GeV, and the associated results from 
EFT are also given for comparison. As expected, the full amplitude calculation lies between the EFT prediction with and without form factor (f.f.), defined as f.f.  $=1/(1+(m_{\gamma\gamma}^{2}/\Lambda')^2)$ with $\Lambda'=1$~TeV. The discrepancy appears because the EFT 
is not valid for such low mass (see Sec.~\ref{se:theory} and Fig.~\ref{fig:mqplane}).  
The backgrounds originate from di-photon and di-electron exclusive production, di-photon and di-jet
production via double Pomeron
exchanges, SM di-photon production with pile up, and SM di-jet and di-electron
production with pile up where the jets or electrons are misidentified as photons.

\begin{figure}
\begin{center}
\includegraphics[scale=0.6]{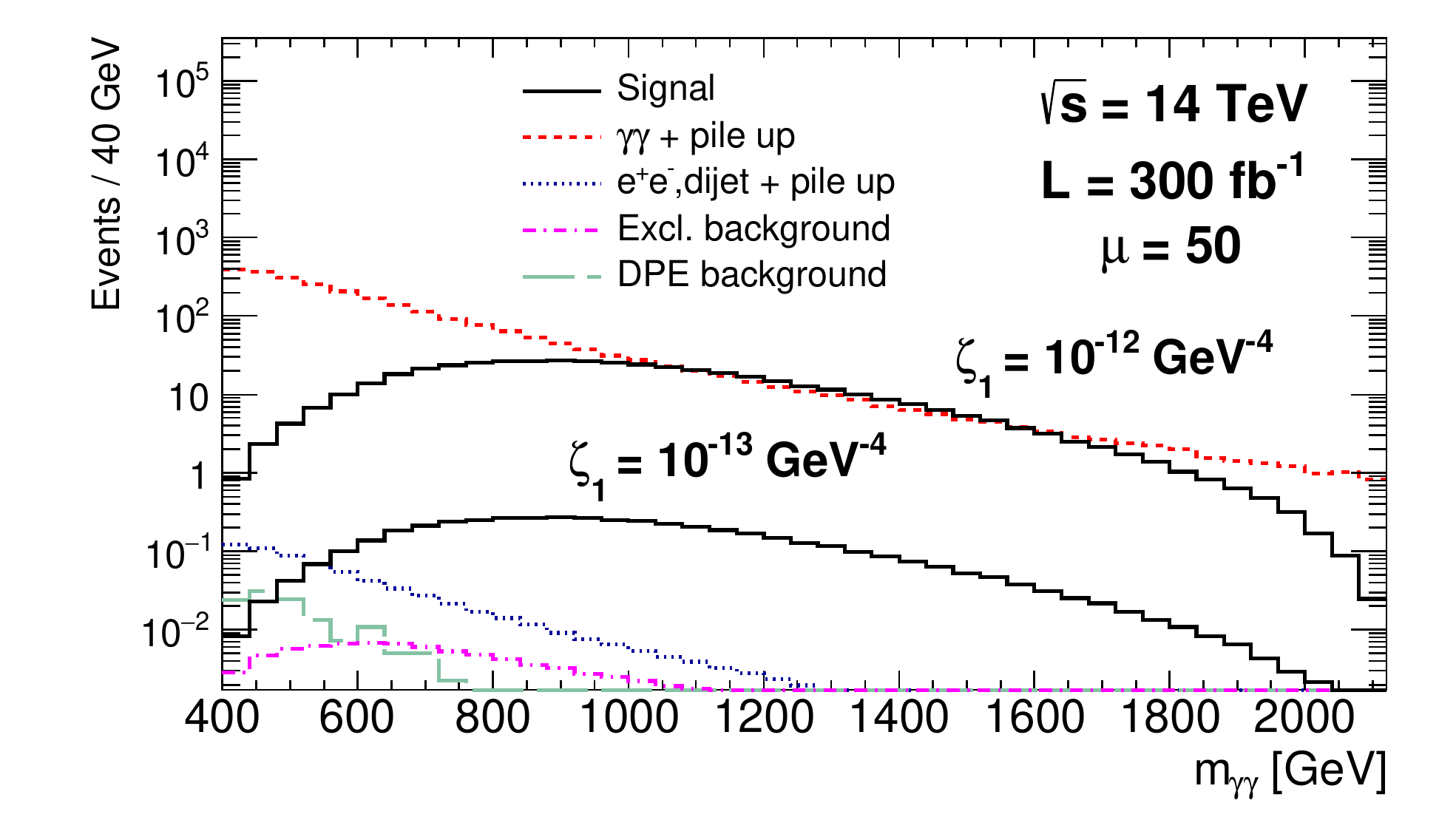}
\end{center}
\caption{ Di-photon invariant
mass distribution for the signal considering two
different coupling values ($10^{-12}$ and $10^{-13}$\gev$^{-4}$, see Eq.~\ref{zetas}) and for the backgrounds (dominated by $\gamma\gamma$ with protons from  pile up), requesting
two intact protons in the forward detectors and two photons in the central detector with a minimun $\pt$ of 200 (100)~GeV for the leading (subleading) photon. The considered integrated luminosity is 
300$~\fbi$ and the pile up $\mu = 50$. Excl. stands for exclusive 
backgrounds and DPE for double pomeron exchange backgrounds (see text).}
\label{fig:mass}
\end{figure}

\begin{figure}
\includegraphics[scale=0.4]{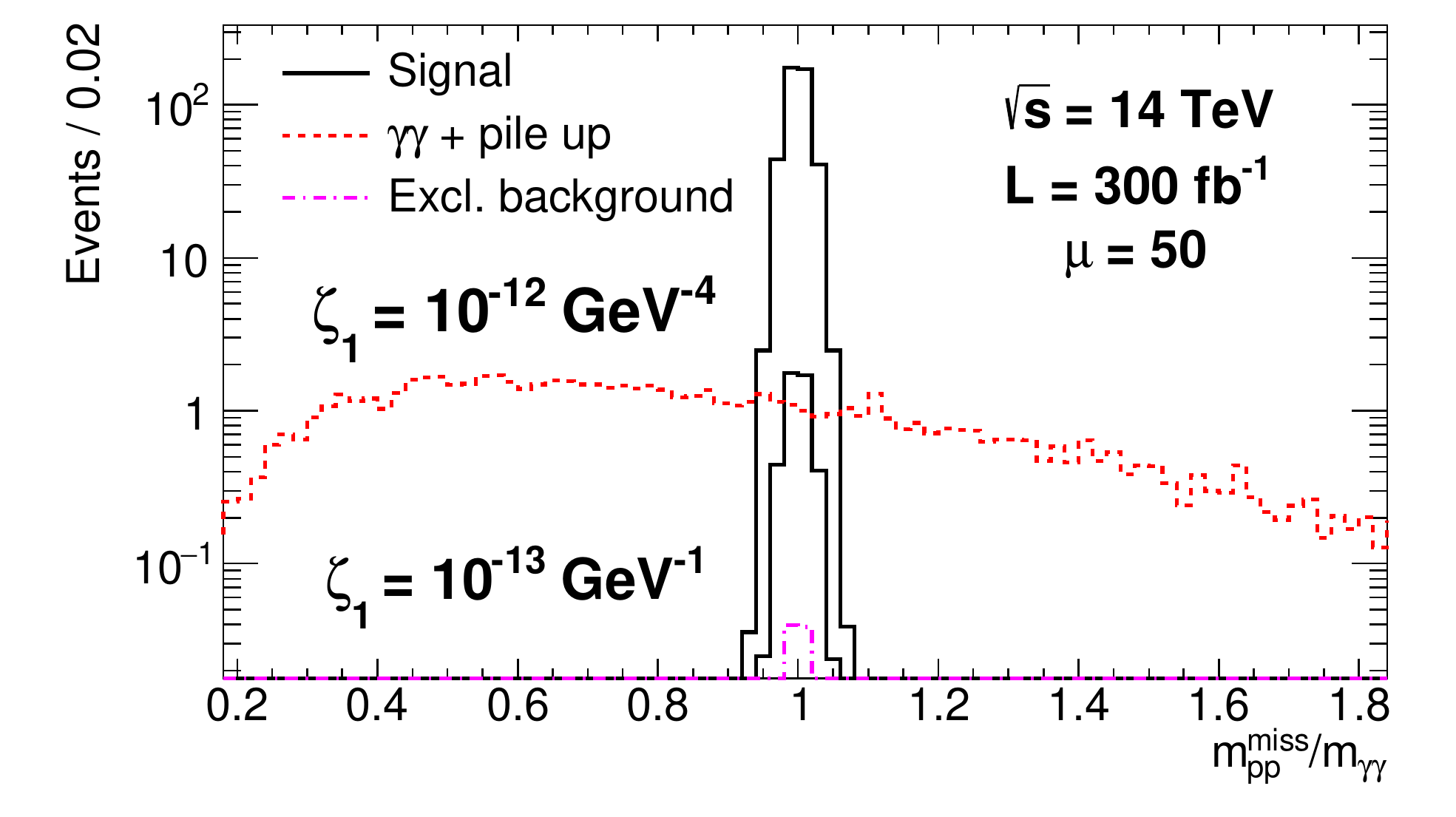}
\includegraphics[scale=0.4]{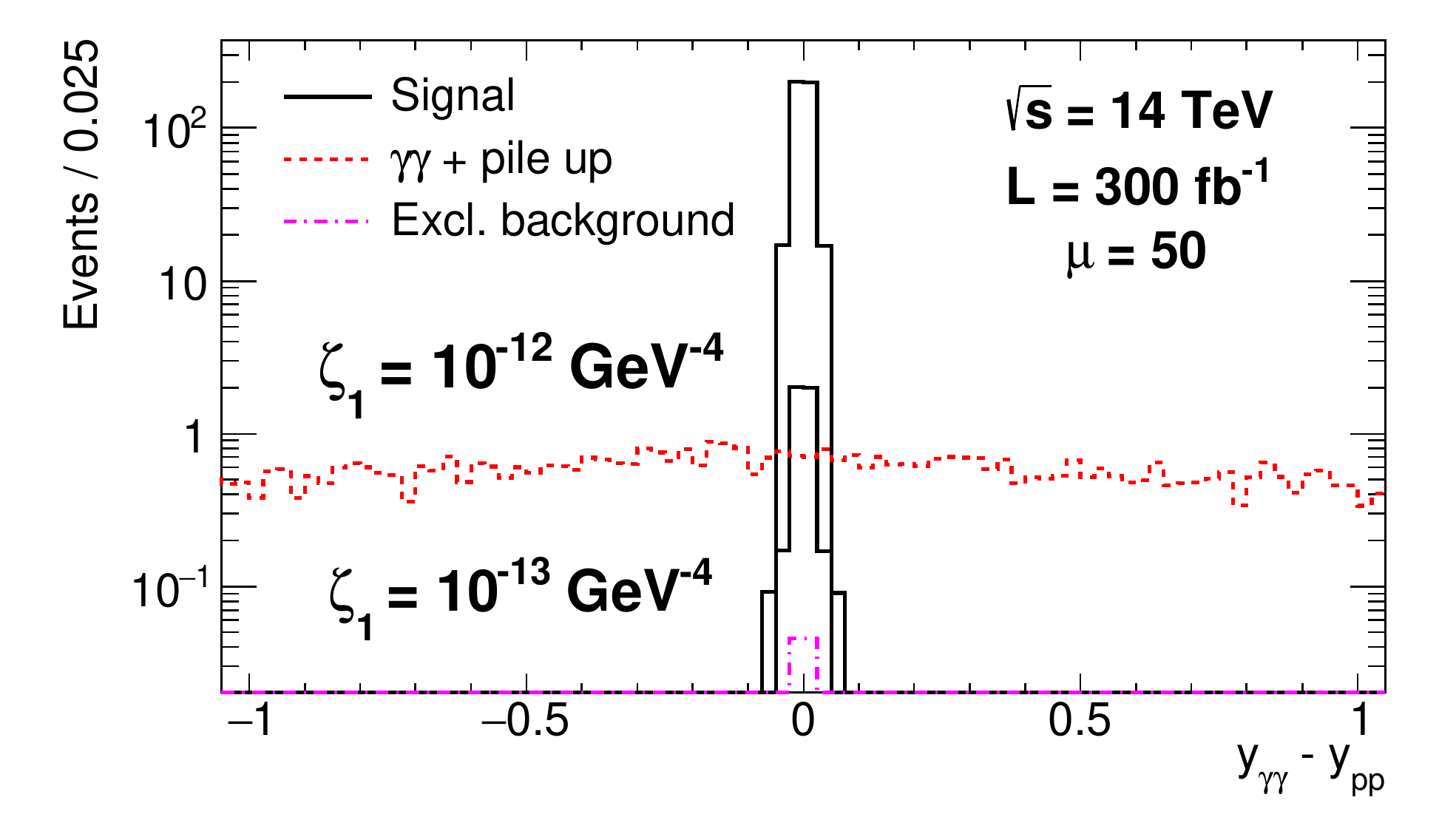}
\caption{\label{fig:massratio} Di-photon to missing proton mass ratio (left) and rapidity difference (right)
distributions for signal considering two 
different coupling values ($10^{-12}$ and $10^{-13}$\gev$^{-4}$, see Eq.~\ref{zetas}) and for 
backgrounds after requirements on photon \pt, di-photon invariant mass, \pt\ ratio between the two photons and on the angle between the two photons. At least one converted photon is required. The integrated luminosity 
is 300~\fbi\ and the average  pile up is $\mu=50$.}
\end{figure}

The different cuts follow the analysis presented in Ref.~\cite{Fichet:2013gsa}. Gaussian 
smearings of 1\% for the total energy, 0.001 for the pseudorapidity and 
1~mrad 
for the azimuthal angle are applied to each photon.
The di-photon mass distribution is given in Fig.~\ref{fig:mass} for the signal
and the different backgrounds. The signal appears at high di-photon masses
whereas the SM background stands at low masses.
The first cuts requires that both protons are measured in AFP or CT-PPS 
($0.015
< \xi< 0.15$) and the photons are produced at high $p_T$ and high mass 
($p_{T1,2}>200, ~100$ GeV and $m_{\gamma \gamma}> 600$ GeV). After those
requirements, the SM exclusive background (dominated by the QED W-loop contribution) is very small (typically 0.1 events). The main
remaining background is the SM di-photon production associated with intact protons from pile up. In order to
suppress this background, the signal event topology is used, requiring that photons are emitted back-to-back and with similar $p_T$. Further
requirements on the exclusivity of the events as shown in Fig.~\ref{fig:massratio} using the forward proton detectors (the di-photon mass and
rapidity are equal within detector resolution when it is computed using the
di-photon in ATLAS/CMS central detector or the proton information from CT-PPS/AFP) completely
suppress the remaining background, while the signal efficiency is over 70\%.
One should notice that tagging the protons
is absolutely fundamental to suppress the $\gamma \gamma$ +  pile up events.
Without the forward proton detector measurements, the number of signal events (64) would be
smaller then the  number of background pile up events (80.2).
Further background reduction is even possible by requiring the photons 
and the protons to originate from the same vertex by measuring their
time-of-flight that provides an additional 
rejection 
factor of 40 for 50  pile up interactions, if one assumes a timing resolution
of 10 ps, showing the large margin on the 
background suppression.
A similar study 
at a higher  pile up of 200 was performed 
and led to a very small background ($<$ 5 
expected background events for 300~\fbi without re-optimizing the event selection), showing the robustness of this 
analysis. 
Moreover, if one relaxes the request of at least one photon to be converted, 
the signal is increased by a factor 3 to 4.
In comparison with the results given in Ref.~\cite{Fichet:2013gsa}, we added the SM di-photon
background induced by $W$ loops and improved the signal generation~\footnote{Our previous estimate was relying on the COMPHEP software~\cite{comphep1,comphep2} interfaced with FPMC.
For this work we implemented directly Eqn.~\ref{xsec} in FPMC. 
We verified that the cross section computed from the full amplitudes reduces to the EFT limit at low energies, providing an independent cross-check of the implementation.
}.

\begin{table}
\small
\begin{tabular}{|c||c|c||c|c|c|c|}
\hline
Cut / Process & \specialcell{Signal\\ (full)} & \specialcell{Signal\\with (without)\\ f.f (EFT) } & Excl. & DPE & \specialcell{DY,\\ di-jet\\+ pile up}  & \specialcell{$\gamma\gamma$\\+ pile up} \\
\hline
\hline
\specialcell{$[0.015<\xi_{1,2}<0.15$, \\ $p_{\mathrm{T}1,(2)}>200,(100)$ GeV]}       & 65   & 18 (187)   & 0.13 & 0.2  & 1.6           & 2968         \\
$m_{\gamma\gamma}>600$~GeV                                                           & 64   & 17 (186)   & 0.10 &  0    & 0.2          & 1023         \\
\specialcell{[$p_{\mathrm{T2}}/p_{\mathrm{T1}}>0.95$,\\ $|\Delta\phi|>\pi-0.01$]}    & 64   & 17 (186)   & 0.10 & 0   & 0         & 80.2         \\
$\sqrt{\xi_{1}\xi_{2}s} = m_{\gamma\gamma} \pm 3\%$                                  & 61   & 16 (175)   & 0.09 & 0   & 0         & 2.8          \\
$|y_{\gamma\gamma}-y_{pp}|<0.03$                                                     & 60   & 12 (169)   & 0.09 & 0   & 0         & 0            \\
\hline
\end{tabular}
\caption{Number of signal events for $S=1$, $Q_{\rm eff}=4$, $m=340$ GeV 
  and background events after 
various selections for an integrated
luminosity of 300~\fbi\ and $\mu=50$ at $\sqrt{s}=14$ TeV.
Values obtained using the corresponding EFT couplings with and without form factors are also displayed.
 At least one 
converted photon is required. Excl. stands for exclusive backgrounds and DPE 
for double pomeron exchange backgrounds (see text).
}
\label{tab:event}
\end{table}

\section{ Expected sensitivity for charged particles at the 14 TeV LHC}
\label{se:lhc}

In this Section, we present the estimates for the LHC sensitivities to new massive charged particles 
obtained through our proposed measurement of light-by-light scattering. 
Results are provided both in the EFT framework and using  full loop amplitudes. Both approaches coincide in the decoupling limit $m\gg E$.

\subsection{EFT results}

\label{se:effective}

We first present the sensitivities for the effective four-photon couplings $\zeta_i$. 
These sensitivities  are given in Table~\ref{sensitivities}
for different scenarios corresponding to the medium luminosity at the LHC
(300 fb$^{-1}$) and the high luminosity (3000 fb$^{-1}$ in ATLAS), with and
without an ad-hoc form factor \cite{Fichet:2013gsa} with a cutoff at 1 TeV. We give the $5 \sigma$ discovery potential as well as the
95\% CL limits with a pile up of 50, and requesting or not at least one photon
to be converted. 

In the table we provide only the sensitivities for ($\zeta_1\neq0$, $\zeta_2=0$) and ($\zeta_1=0$, $\zeta_2\neq0$).
It turns out that with the cuts adopted in our analysis (see Sec.~\ref{se:bg}), the interference of the $\zeta_i$ vertices with the background is negligible. The cross section has thus the form 
given in Eq.~(\ref{xsec}) to a good approximation. Knowing the sensitivity for a given value of $\zeta_1$, $\zeta_2$, it is straightforward to recast it in the complete  $(\zeta_1,\zeta_2)$ plane, where it defines an ellipse 
\be
\zeta^2\equiv\ 48 (\zeta_1)^2 + 40 \zeta_1 \zeta_2 + 11 (\zeta_2)^2\,.
\ee
\label{zetaeff}
We find that the sensitivity extends up to $|\zeta| \approx 7\cdot10^{-14}$ GeV$^{-4}$. 

 Using a form factor with higher cutoff $\gtrsim 2$ 
TeV leads to similar results as without form factors. The reach is slightly
better than in our previous study~\cite{Fichet:2013gsa} because of the increase of the
number of signal events (see footnote \textcolor{blue}{2}). The obvious inconvenience of the EFT approach is that it is valid only in the high mass region, $m\gg E$. In order to use the EFT result down to $m\sim E$, it is common to introduce ad-hoc form factors which mimics the behaviour of the -- unknown -- amplitudes near the threshold. 
Clearly, this method introduces a great deal of arbitrariness into the results. Not only do the results depend on the functional form of the form factor, but also on the energy scale at which they are introduced.

\begin{table}

\begin{center}
\begin{tabular}{|c||c|c||c|c||c|}
\hline
Luminosity & 300~\fbi & 300~\fbi & 300~\fbi & 300~\fbi & 3000~\fbi \\
\hline
 pile up ($\mu$) & 50 & 50 & 50 & 50 & 200 \\
\hline
\hline
coupling & $\ge$~1 conv. $\gamma$ & $\ge$~1 conv. $\gamma$ & all $\gamma$ & all $\gamma$  & all $\gamma$\\
(GeV$^{-4}$) & 5 $\sigma$ & 95\% CL & 5 $\sigma$ & 95\% CL & 95\% CL \\

\hline
$\zeta_1$~f.f.   &  $1.5\cdot10^{-13}$   & $7.5\cdot10^{-14}$   & $6\cdot 10^{-14}$ & $4\cdot 10^{-14}$ & $3.5\cdot10^{-14}$ \\
$\zeta_1$~no f.f.&  $3.5\cdot10^{-14}$ & $2.5\cdot10^{-14}$     & $2\cdot10^{-14}$  & $1\cdot10^{-14}$  & $1\cdot10^{-14}$\\
\hline
$\zeta_2$~f.f.   &  $2.5\cdot10^{-13}$   & $1.5\cdot10^{-13}$   & $1.5\cdot10^{-13}$    & $8.5\cdot10^{-14}$  & $7\cdot10^{-14}$ \\
$\zeta_2$~no f.f.&  $7.5\cdot10^{-14}$   & $4.5\cdot10^{-14}$   & $4\cdot10^{-14}$&    $2.5\cdot10^{-14}$  & $2.5\cdot10^{-14}$ \\
\hline

\end{tabular}
\end{center}

\caption{5\,$\sigma$ discovery and 95\% CL exclusion limits on $\zeta_1$ and $\zeta_2$ 
couplings in\gev$^{-4}$ (see Eq.~\ref{zetas}) with
and without form factor (f.f.), requesting at least one converted photon ($\ge$~1 conv. $\gamma$) or
not (all $\gamma$). All sensitivities are given for 300 fb$^{-1}$
and $\mu=50$  pile up events (medium luminosity LHC) except for the numbers of the last column which are given for 3000
fb$^{-1}$ and $\mu=200$  pile up events (high luminosity LHC). }
\label{sensitivities}

\end{table}

\subsection{Results from exact amplitudes}

This section contains our more general results for charged particles. Contrary to the EFT approach which is valid on in the decoupling limit and requires ad-hoc form factors to be valid near the threshold, the use of the full amplitudes provides exact results  for any mass.

The results are given in Tab. \ref{fullamp_values} and
Fig.~\ref{fig:mqplane}  where we display the 5$\sigma$ discovery, 3$\sigma$
evidence and 95\% C.L.~limit for fermions and vectors 
for a luminosity of 300 fb$^{-1}$ and a pile up of 50. 
We find that a vector (fermion) with  $Q_{\rm eff}=4$, can be discovered up to mass $m=640$~GeV ($300$~GeV). At high mass, the exclusion bounds follow isolines $Q\propto m$, as dictated by the EFT couplings Eq. \ref{zetaEFT}. Extrapolating the same analysis to a higher luminosity of 3000 \fbi for a pile up of 200 leads to a slighlty improved sensitivity of  $m = 680$~GeV ($340$~GeV) for vectors (fermions).

Comparing with our discussion in Sec.~\ref{se:general},  one notices that some searches for VL quarks, as motivated from e.g.~Composite Higgs models, already lead to stronger bounds than the ones projected here. For instance, VL top partners arising from the $(2,2)$ (corresponding to $Q_{\rm eff}\approx2.2$) of mass $m=500$ GeV would be excluded from present LHC data, while they would be out of reach in our method.
On the other hand, our results are completely model-independent. They apply 
just as well to different effective charges, are independent of the amount 
of mixing with the SM quarks, and even apply to VL leptons.  

\subsection{Comparison with the muon $g-2$ measurement}

\begin{figure}
\begin{center}
\includegraphics[width=0.49\linewidth]{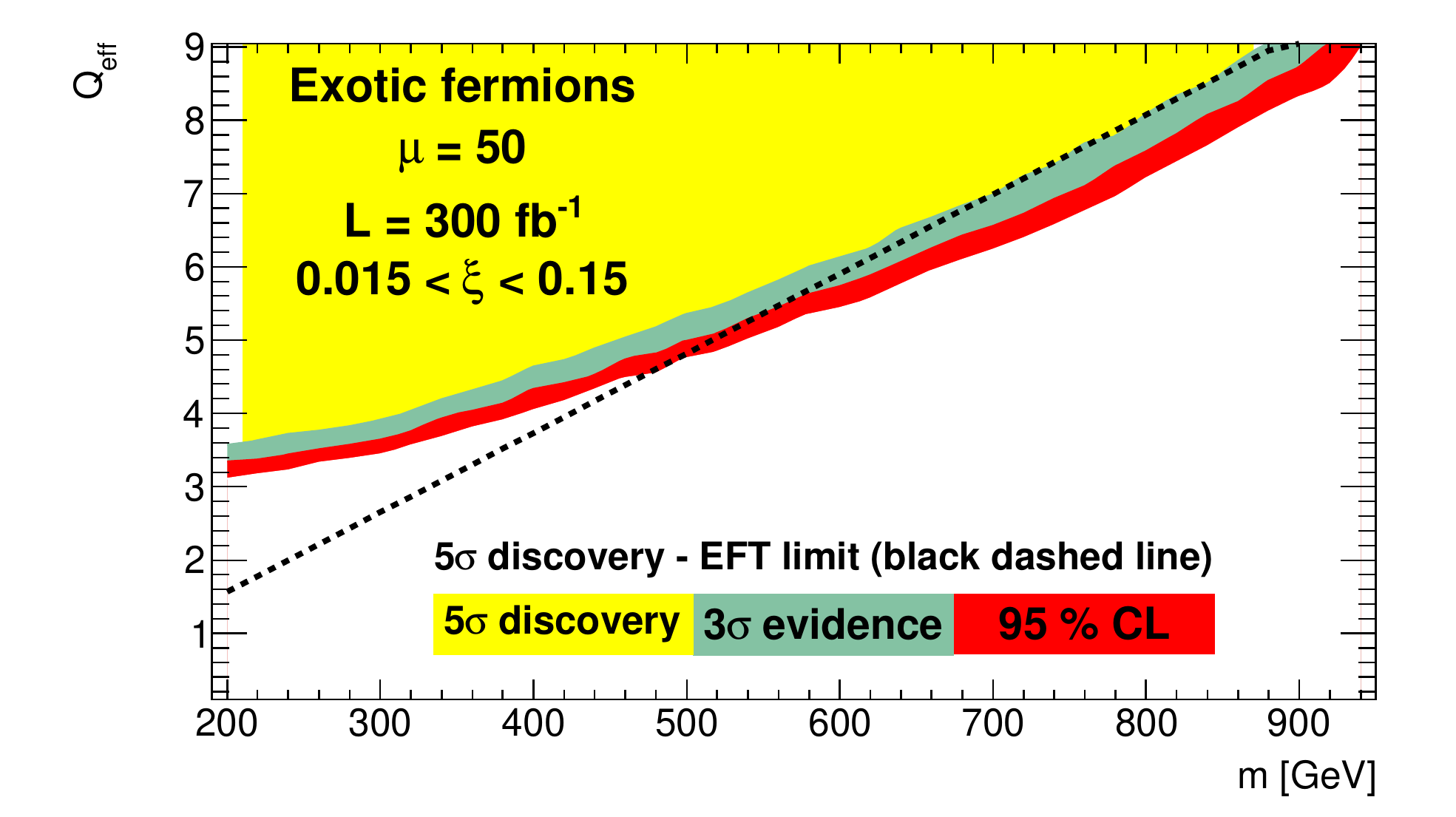}
\includegraphics[width=0.49\linewidth]{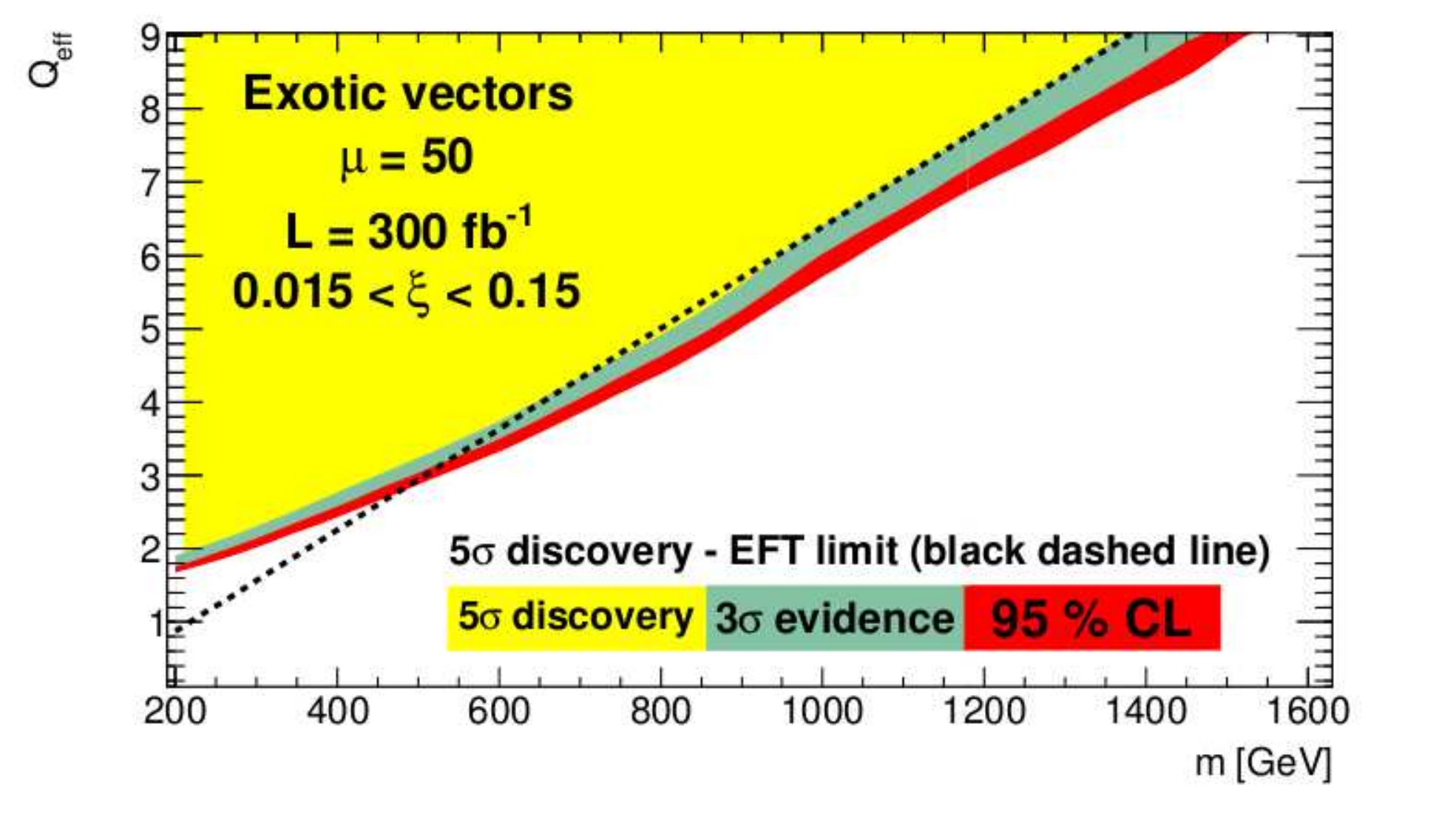}
\end{center}
\caption{Exclusion plane in terms of mass and effective charge of generic fermions and vectors. in the case of no requirement of photon conversion at the analysis stage and full integrated luminosity at the medium-luminosity LHC (300~\fbi, $\mu=50$).}
\label{fig:mqplane}
\end{figure}

\begin{figure}

\begin{center}
\includegraphics[scale=.66]{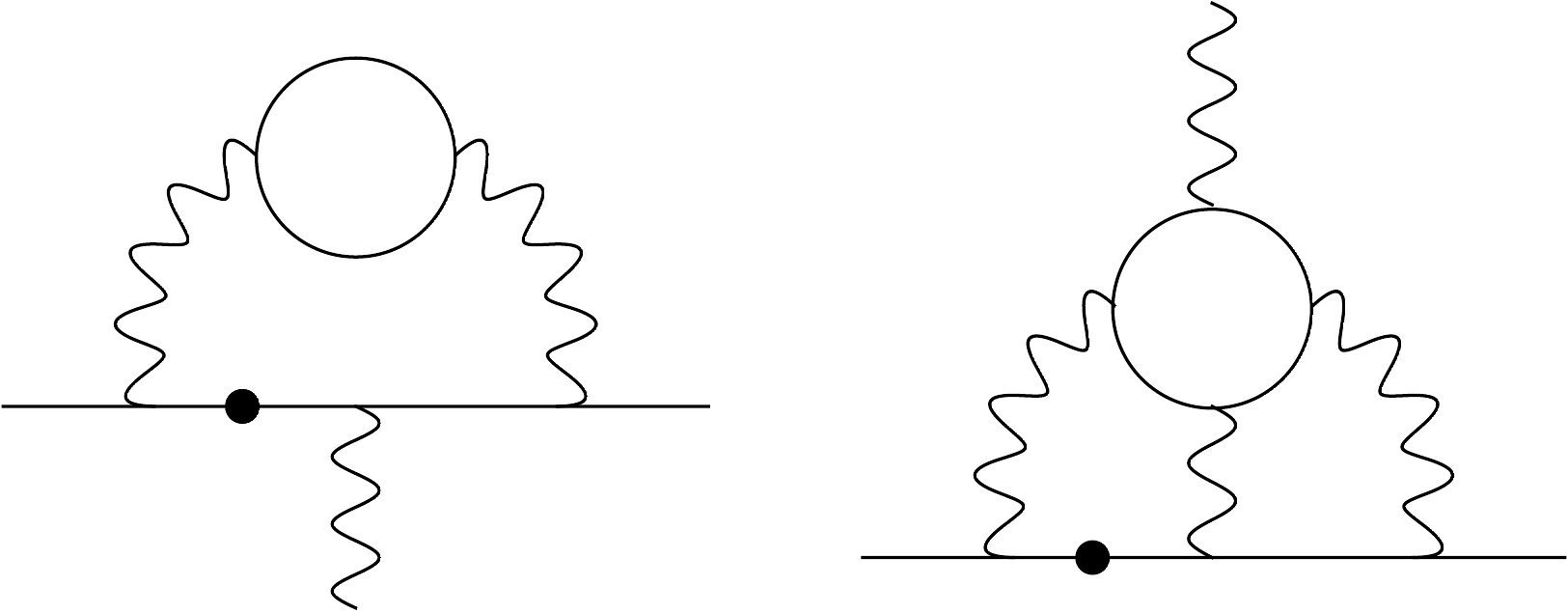} 
\end{center}
\caption{\label{fig:g-2}
Two and three loop contribution to the muon anomalous gyromagnetic factor. The dot represents a muon mass insertion and the circle a generic NP particle of mass $m$ and charge $Q_{\rm eff}$.
}
\end{figure}

Finally we would like to comment on the possibility to observe charged particles in other precision observables. For instance, they could contribute to the magnetic dipole moment of the muon via a higher-loop diagrams as shown in Fig.~\ref{fig:g-2}.\footnote{We would like to thank O.~Lebedev for suggesting this possibility.} The contribution to the coefficient of the effective dipole operator $F_{\mu\nu}\bar\mu_L\sigma^{\mu\nu}\mu_R$, up to $\mathcal O(1)$ numbers and factors of $\log (m_\mu/m)$ can be estimated as~\footnote{Strictly speaking, the two-loop graph is proportional to $\tr Q^2$ and not to $Q_{\rm eff}^2=(\tr Q^4)^\frac{1}{2}$. We ignore this small difference as we are content with a rough estimate here.}
\be
d^{(2-{\rm loop})}_\mu\sim \frac{e^5Q_{\rm eff}^{2} m_\mu}{m^2(16 \pi^2)^2}\,,\qquad 
d^{(3-{\rm loop})}_\mu\sim \frac{e^7Q_{\rm eff}^{4} m_\mu}{m^2(16 \pi^2)^3}
\ee 
leading to contributions to $a_\mu=4\,d_\mu m_\mu/e$ of the order of
\bea
\Delta a_\mu^{(2)} %\sim \frac{4e^4Q_{\rm eff}^2 m_\mu^2}{m^2(16 \pi^2)^2
&\sim&1.5\cdot 10^{-12}\left(\frac{100 {\rm \ GeV}}{m}\right)^2 Q_{\rm eff}^2\,,\nn\\
\Delta a_\mu^{(3)} %\sim \frac{4e^6Q_{\rm eff}^4 m_\mu^2}{m^2(16 \pi^2)^3}
&\sim&8.6\cdot 10^{-16}\left(\frac{100 {\rm \ GeV}}{m}\right)^2 Q_{\rm eff}^4\,.
\eea
Unless the effective charge is extremely large we can focus on just the two-loop contribution.
The current experimental uncertainty for $a_\mu$ is around $\sim 6\cdot 10^{-10}$, implying a sensitivity of this measurement to $m/Q_{\rm eff}\sim 5$ GeV. 
Comparing this estimate to our projections from Fig.~\ref{fig:mqplane} we see 
that, despite its impressive accuracy, the $g-2$ measurement is not 
competitive with our method. 

\begin{table}

\begin{center}
\begin{tabular}{|c||c|c|c|c|}
\hline
Mass (GeV) & 300 & 600 & 900 & 1200 \\
\hline
$Q_{\rm eff}$ (vector)  & 2.3 & 3.7 & 5.6 & 7.8 \\
\hline
$Q_{\rm eff}$ (fermion) & 3.9 & 6.2 & 9.1 & - \\
\hline
\end{tabular}
\end{center}

\caption{5\,$\sigma$ discovery limits on the effective charge of new generic charged fermions and vectors for various masses scenarios in the case of no requirement of photon conversion at the analysis stage and full integrated luminosity at the medium-luminosity LHC (300~\fbi, $\mu=50$).}
\label{fullamp_values}

\end{table}

\section{Neutral contributions to light-by-light scattering \label{se:otherNP} }

\begin{figure}
\begin{picture}(400,250)
\put(50,0){		\includegraphics[trim=0cm 0cm 0cm 0cm, clip=true,width=10cm]{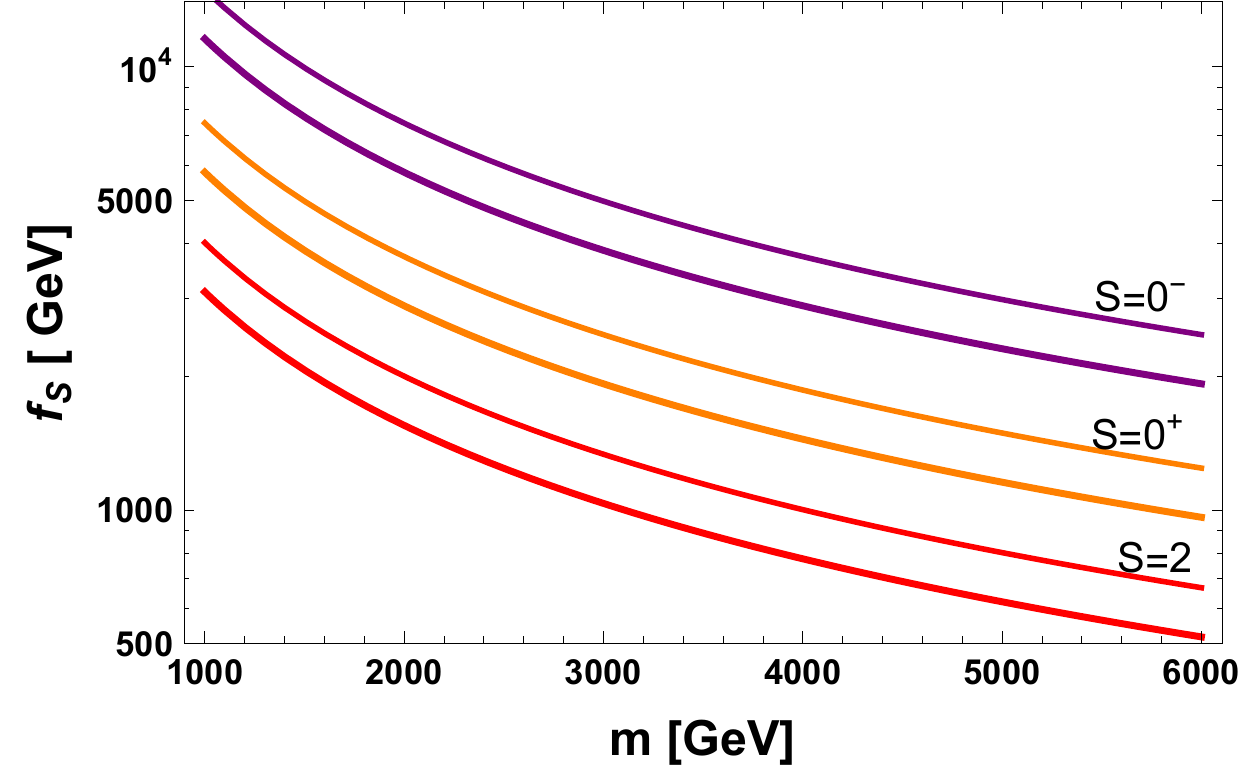}}
\end{picture}
\vspace{0.5cm}
	\caption{ Sensitivities for the neutral simplified models in the $(m,f_S)$ plane. 	
	Thick lines correspond to 5\,$\sigma$, thin lines correspond to 95\% CL limits. 
The limits are given for the medium luminosity LHC with all photons (no conversion required) and no form-factor (see Tab. \ref{sensitivities}).		
	 \label{fig:fm_plot}}
	\end{figure}

In the decoupling limit, any new physics contribution can be mapped onto the $(\zeta_1,\zeta_2)$ plane. 
This provides a model-independent way to compare and summarize the discovery reach for various candidates. 
The EFT mapping for charged particles has been discussed in Sec. \ref{sec:eft}. 
Beyond perturbative contributions to $\zeta_i^\gamma$ from  
charged particles,  non-renormalizable interactions of neutral particles are 
also present in common extensions of the SM.  Such theories can contain 
scalar, pseudo-scalar and spin-2 resonances, respectively denoted by $\varphi$, 
$\tilde \varphi$ and $h^{\mu\nu}$. Independently of the particular New Physics model they originate from, their leading couplings to the photon  
are fixed completely by Lorentz and CP symmetry as
\be\begin{split}
\mathcal L_{\gamma\gamma}=&f_{0^+}^{-1}\,\varphi\, 
(F_{\mu\nu})^2+f_{0^-}^{-1}\, \tilde\varphi \, F_{\mu\nu}F_{\rho\lambda}\,
\epsilon^{\mu\nu\rho\lambda} \\&+f_{2}^{-1}\, h^{\mu\nu}\, (-F_{\mu\rho} 
F_{\nu}^{\,\,\rho}+\eta_{\mu\nu} (F_{\rho\lambda})^2/4)\,,
\end{split}
\ee
where the $f_S$ have mass dimension 2. 
They then generate  $4\gamma$ couplings by tree-level exchange as 
$\zeta_i=(f_{S}\, m)^{-2}\,d_{i, s}$, where
\be
d_{1,s}=
\begin{cases}
\frac{1}{2} & s=0^+ \\
 -4 & s=0^- \\
-\frac{1}{8} & s=2\\
\end{cases}
\,,\quad
d_{2,s}=
\begin{cases}
0 & s=0^+ \\
8  & s=0^- \\
\frac{1}{2} & s=2 \\
\end{cases}
\,.
\ee
We show in Fig.~\ref{fig:fm_plot} our model independent sensitivities for these three cases.

\begin{figure}
\begin{picture}(400,250)
\put(50,0){		\includegraphics[trim=0cm 0cm 0cm 0cm, clip=true,width=10cm]{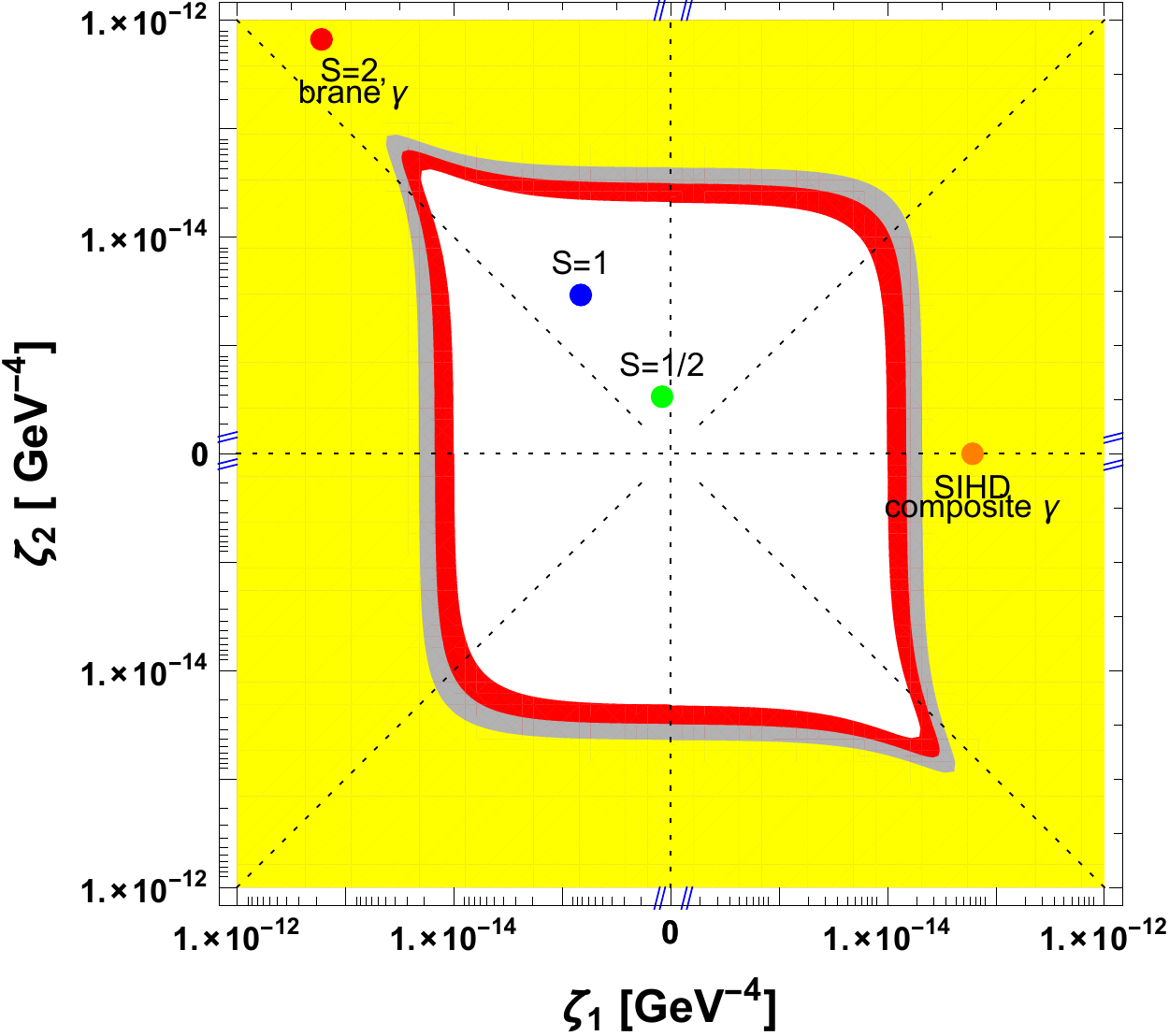}}
\end{picture}
\vspace{0.5cm}
	\caption{ Experimental sensitivity and models in the $(\zeta_1,\zeta_2)$ plane. Axes follow a logarithmic scale spanning $|\zeta_i|\in[10^{-12},10^{-16}]$.
	%\cup {0} \cup [10^{-12},10^{-16}]\,\textrm{GeV}^{-4}$. 
	The yellow, grey, and red  regions can be probed at 5\,$\sigma$, 3\,$\sigma$ and 95\%~CL using proton tagging at the LHC, while 
	the white region remains inaccessible. 	The limits are given for the medium luminosity LHC with all photons (no conversion required) and no form-factor (see Tab. \ref{sensitivities}).	
	Also shown are contributions from electric particles with spin $1/2$ 
	and $1$, charge $Q_{\rm eff}=3$, mass $m=1$ TeV,  
	the contribution from warped KK gravitons with mass $m_{\rm KK}=3$~TeV, 
	$\kappa=2$ and brane-localized photon, and the
	contribution from a strongly-interacting heavy dilaton (SIHD) 
	with mass $m_{\varphi}=3$ TeV coupled to a composite photon.
	 \label{fig:zeta_plot}}
	\end{figure}

Various contributions from New Physics states are shown together with the discovery reach estimated in this paper 
 in Fig.~\ref{fig:zeta_plot}. We stress that this plot is valid 
 only in the decoupling limit, so that the results based on full amplitudes 
 presented in Sec. \ref{se:lhc} are not included in the plot. 

Let us review the other known new physics candidates for completeness (see \cite{Fichet:2013gsa,Fichet:2013ola} for complementary information).

\begin{itemize}

\item {\it Kaluza-Klein gravitons:}
Kaluza Klein gravitons of warped extra dimensions also produce effective four-photon vertices. The contribution of the entire tower of resonances, in the case that the SM resides on the Infrared brane, leads to \cite{Fichet:2013ola}
\be
\zeta_1=-\frac{\kappa^2}{64\tilde k^4}\,,\qquad
\zeta_2=\frac{\kappa^2}{16\tilde k^4}
\ee
where $\tilde k$ is related to the extra dimensional curvature and sets the mass of the KK resonances as $m_{\rm KK}\approx 3.8 \tilde k$. The quantity $\kappa$ sets the coupling strength and can be taken of order unity.
For $\kappa=2$, and using the 5\,$\sigma$ and 95\% CL sensitivities   for the medium luminosity LHC with all $\gamma$s and no form-factor (see Tab. \ref{sensitivities}), the effect of the KK resonances can be detected up to  mass 
\be
m_{\rm KK}<5240\, \textrm{GeV} \,(5\,\sigma)\,,\quad m_{\rm KK}<6230\, \textrm{GeV}\,\,(95\%{\rm CL})\,.
\ee	
These sensitivities are competitive with respect to searches for direct production of KK resonances at the LHC.

\item{\it Strongly-interacting dilaton:}
Extensions of the Standard Model sometimes feature a new strongly-interacting sector. Provided that this sector is conformal in the UV, it is most likely explicitly broken in the IR, at least by the appearance of electroweak scale and QCD confinement. As a result, the spectrum of the strong sector features a neutral scalar, the so-called dilaton, whose mass lies close to the scale of conformal breaking. In the absence of fine-tuning the dilaton's couplings are unsuppressed with respect to this scale. To distinguish it from the weakly coupled (fine-tuned) light dilaton often considered in the literature we will refer to it as the Strongly-Interacting Heavy Dilaton (SIHD). If the photon is at least partially composite, it also couples strongly to the dilaton. 
For a pure composite photon, one gets
\be
\zeta_1=\frac{\pi^2}{2m_\varphi^4}\,,\qquad
\zeta_2=0\,.
\ee
By the AdS/CFT correspondence, the SIHD is equivalent to the radion in warped extradimension scenarios where the metric presents sizeable departure from $AdS_5$ in the IR. The photon can couple strongly if it is either localized on the IR brane or if it has a sizable IR brane kinetic term.
Using the 5\,$\sigma$ and 95\% CL sensitivities for the medium luminosity LHC with all $\gamma$s and no form-factor (see Tab. \ref{sensitivities}), the effect of the SIHD can be detected up to mass
\be
m_{\varphi}<3960\, \textrm{GeV}\,(5\,\sigma)\,,\quad m_{\varphi}<4710\, \textrm{GeV}\, (95\%{\rm CL})\,.
\ee	
  
\end{itemize}

One should notice that the expected observation of four-photon interactions 
at the LHC discussed in this work does not provide information 
on $\zeta_1$, $\zeta_2$ separately, nor their sign, but only on the 
combination Eq.~(\ref{zetaeff}). In principle, more refined observables 
could provide information lifting this degeneracy. 
 This would in turn provide a way of identifying various contributions. 
As the various amplitudes depend on different combinations of $\zeta_1$ and $\zeta_2$,  polarization-based observables could play this role. One can notice, for example, that $\mathcal{M}_{++--}$ is exactly zero for a spin-2 particle as the  KK graviton. 
However, in the measurement proposed in this paper, we find that no information discriminating between $\zeta_1$, $\zeta_2$ can be obtained by looking at the photon  angular distributions.

\section{Conclusion}

The scheduled installation of forward proton detectors at the LHC will provide a -- somewhat surprising -- opportunity to measure the scattering of light-by-light, providing a new window on physics beyond the Standard Model. 
This paper is dedicated to the estimation of the discovery potential for light-by-light scattering at the 14 TeV LHC, especially in the case of generic new electrically charged particle of spins $1/2$ and $1$.  
 
Light enough  charged particles could in principle be directly produced and 
observed at the LHC. However such processes are highly model-dependent, and a 
dedicated analysis has to be set up for each specific case. In contrast, light-by-light scattering provides model-independent limits from a single precision measurement, such that both approaches are  complementary.

A former estimation of the LHC sensitivity to heavy charged particles has been performed in \cite{Fichet:2013gsa} in the decoupling limit. The case of light masses \ie~lower than a few TeV, which is potentially the most interesting, is not covered in this approach, as it requires the use of ad-hoc form-factors. To avoid the introduction of such arbitrariness, we implemented the full one-loop amplitudes for spin $1/2$ and spin $1$, such that our simulations are valid for any mass.

The crucial feature of the forward proton detectors is that they give access to the complete kinematics of the events. This can be used to reject most of the background.
The implementation of the  generic one-loop amplitudes contributions -- including all limiting regimes -- is done in FPMC.
The implementation of the full amplitudes is also useful to simulate the SM QED background. It appears that the W-loop  dominates over fermion loops, and takes a simple form in the  high-energy regime.

We provide the sensitivity to charged particles in the $(Q_{\rm eff},m)$ plane for medium and high luminosity scenarios. 
For $Q_{\rm eff}=4$, we find that a new vector (fermion) can be detected at 5\,$\sigma$  up to mass $m=640$~GeV ($300$~GeV)  
and $m=680$~GeV ($340$~GeV)  respectively for the medium and high luminosity 
LHC configuration. The transition between EFT and full amplitudes results is also discussed quantitatively.

We  also point out that  new charged particles  contribute to the muon anomalous gyromagnetic moment at two and three-loop. An estimate of these contributions shows that, in spite of the  impressive precision of the muon g-2 measurement, it cannot compete with the LHC search we propose. 

The inclusive tri-photon cross section is also sensitive to the anomalous four-photon couplings since it can be
produced by annihilation of a quark-antiquark pair into a photon which then splits into 
three photons. 
This channel will be studied in detail at the LHC, but we expect its sensitivity not to be as good as the reach obtained in this paper. Indeed, three photons are 
produced in the final state so the transverse momentum of the third photon is 
smaller, leading to a more complicated analysis.

The light-by-light scattering sensitivity to neutral particles in the EFT limit is also considered  through simplified models. We find that 
warped KK gravitons and  the strongly-interacting heavy dilaton can be discovered in the multi-TeV range.

Finally, looking at the $s=0,1/2,1$ charged particles contributions (Eq.~\eqref{zetaEFT}, \eqref{EH}), we notice that the contributions from charged loops appear to grow quite fast with the spin of the particle. If this behaviour remains true for larger spin, light-by-light scattering might constitute an interesting probe for the presence of higher-spin particles, like string excitations or strongly-interacting bound states. Further tools are however necessary to handle quantum computations involving higher-spin particles and are under developments  \cite{fgwip}. 

\section*{Acknowledgements} 

We would like to thank O.~Kepka and O.~Lebedev for valuable discussions.
SF acknowledges the Brazilian Ministry of Science, Technology and Innovation for financial support, and the Institute of Theoretical Physics of Sao Paulo (ICTP/SAIFR) for hospitality. GG would like to thank the Funda\c c\~ao de Amparo \`a Pesquisa do Estado de S\~ao Paulo (FAPESP) for financial support.
\\
\\
\\

\noindent{\Large\bf Appendix}

\appendix

\section{The full one-loop amplitudes}
\label{app:Amplitudes}

In this appendix we collect for completeness the expressions for the on-shell four-photon amplitudes generated from heavy spin-$1/2$ and spin-1 states \cite{Costantini:1971cj,Jikia:1993tc}. We also compute various kinematical limits. These expressions can be applied to both SM particles (quarks, leptons, $W-$ bosons, as well as any New Physics particles of spin $\frac{1}{2}$ and $1$. In particular, for the SM contributions, we need to sum over the quark and lepton spectrum with the correct electric charges ($-1$ for leptons, $\frac{2}{3}$ for up type quarks, $-\frac{1}{3}$ for down-type quarks and $1$ for the $W$-boson.)

\subsection{Loop functions}

The loop-integrals can be expressed in terms of the functions $B(z)$, $T(z)$ and $I(z,w)$, defined as
\bea
\Re B(z)&=&-1+\Re \left[\frac{b(z)}{2}\log\left(\frac{b(z)+1}{b(z)-1}\right)\right]\nn\\
\Im B(z)&=&-\frac{\pi}{2}b(z)\qquad {\rm for\ }z>1\,.
\eea
where $b(z)=\sqrt{1-1/z}$ as well as
\bea
\Re T(z)&=&\Re \left[\frac{1}{4}\log^2\left( \frac{b(z)+1}{b(z)-1}  \right)\right]\nn\\
\Im T(z)&=&-\pi \arcosh\sqrt{z}\qquad {\rm for\ }z>1\,.
\eea
and

\bea
\Re I(z,w)&=&\frac{1}{2a}\Re\left[ %\log [z(a^2-b^2(z))]\log\left(\frac{a+1}{a-1}\right) 
-\Li\left(\frac{a+1}{a+b(z)}\right)
+\Li\left(\frac{a-1}{a+b(z)}\right)
-\Li\left(\frac{a+1}{a-b(z)}\right)
+\Li\left(\frac{a-1}{a-b(z)}\right)
\right.\nn\\
&& \left.    
-\Li\left(\frac{a+1}{a+b(w)}\right)
+\Li\left(\frac{a-1}{a+b(w)}\right)
-\Li\left(\frac{a+1}{a-b(w)}\right)
+\Li\left(\frac{a-1}{a-b(w)}\right)
\right]\nn\\
\Im I(z,w)&=&\frac{\pi}{2a}\left[ \Theta(z-1)\log\left( \frac{a-b(z)}{a+b(z)}\right) +\Theta(w-1)\log\left(\frac{a-b(w)}{a+b(w)} \right)\right]\,,
\eea
where $a(z,w)=\sqrt{1-1/z-1/w}$, $\Li(z)=-\int_0^z\log(1-t)/t$ is the dilogarithm function, and $\Theta(x)$ is the units step function that is 0 (1) for $x<0$ ($x>0$). 

\subsection{Amplitudes}

It proves useful to define the rescaled Mandelstam variables
\be
s'=\frac{s}{4 m^2},\qquad t'=\frac{t}{4m^2}\,\qquad u'=\frac{u}{4m^2}
\ee 
where $s'+t'+u'=0\,,$
and $-s'\leq t,u\leq 0$. 
Here, $m$ denotes the mass of the particle in the loop. 
The helicity amplitudes for fermion loops have been computed in Ref.~\cite{Costantini:1971cj}, %\footnote{
%The translation of the notation of Ref.~\cite{Costantini:1971cj} to standard notation reads
%\be
%\mathcal M^{\rm std}_{\lambda_1\lambda_2\lambda_3\lambda_4}
%= -8i\alpha^4\mathcal M^{\rm CTP}_{-\lambda_1-\lambda_2\lambda_3\lambda_4}
%\ee
%}
they are given as
%\begin{widetext}
\bea
\mathcal M^f_{++++}&=&1+
2\left[\frac{t'-u'}{s'}\right]\left[B(t')-B(u')\right]
%+\left[2+\frac{4u'}{s'}\right]B(u')
+\left[\frac{2(t'^2+u'^2)}{s'^2}-\frac{2}{s'}\right]\left[T(t')+T(u')\right]\nn\\
&& +\left[\frac{1}{2s't'}-\frac{1}{t'}\right]I(s',t')
   +\left[\frac{1}{2s'u'}-\frac{1}{u'}\right]I(s',u')\nn\\&&
   +\left[\frac{4}{s'}+\frac{1}{t'}+\frac{1}{u'}+\frac{1}{2t'u'}-\frac{2(t'^2+u'^2)}{s'^2}\right]I(t',u')\nn\\
\mathcal M^f_{+++-}&=&-1-\left[\frac{1}{s'}+\frac{1}{t'}+\frac{1}{u'}\right]\left[T(s')+ T(t')+T(u') \right]\nn\\&&
+\left[\frac{1}{u'}+\frac{1}{2s't'}\right]I(s',t')
+\left[\frac{1}{t'}+\frac{1}{2s'u'}\right]I(s',u')
+\left[\frac{1}{s'}+\frac{1}{2t'u'}\right]I(t',u')
\nn\\
\mathcal M^f_{++--}&=&-1
+\frac{1}{2s't'}I(s',t')
+\frac{1}{2s'u'}I(s',u')
+\frac{1}{2t'u'}I(t',u')
\eea

The helicity amplitudes for vector loops taken from \cite{Jikia:1993tc} read
\bea
\mathcal M^v_{++++}&= & 
-\frac{3}{2} - 3\left[\frac{t'-u'}{s'}\right]\left[B(t')-B(u')\right]
 -   \frac{1}{s'} \left[8 s' - 3 - 6  \frac{t' u'}{s'}\right] \left[T(t') + T(u') - I(t', u') \right] 
\nn \\&&  - \frac{3}{s'} I(t', u') -   4 (s' - \frac{1}{4}) (s' - \frac{3}{4}) \left[ I(s', t')\frac{1}{s' t'} + I(s', u')\frac{1}{s' u'} + I(t', u')\frac{1}{t' u'} \right] 
\eea
whereas the other two are simply rescaled w.r.t.~the fermion result, 
$\mathcal M^v_{+++-}=  -\frac{3}{2} M^f_{+++-}$ and 
$\mathcal M^v_{++--}=  -\frac{3}{2} M^f_{++--}$.

%\end{widetext}

\section{Limits}

\subsection{Low-Energy Approximation}

In the low-energy limit $s',|t'|,|u'|\ll 1$ the amplitudes become
\be
\mathcal M^f_{++++}=-\frac{22}{45}s'^2\,,\qquad
\mathcal M^f_{++--}=\frac{2}{15}(s'^2+t'^2+u'^2)\,,\qquad
\mathcal M^f_{+++-}=\mathcal O(s'^3)\,.
\label{LEf}
\ee
for the fermion loops and
\be
\mathcal M^v_{++++}=-\frac{28}{5}s'^2\,,\qquad
\mathcal M^v_{+++-}=\frac{1}{5}(s'^2+t'^2+u'^2)\,,\qquad
\mathcal M^v_{++--}=\mathcal O(s'^3)\,.
\label{LEv}
\ee
for the vector loops.

\subsection{High-Energy Approximation}

In the high-energy limit $s',|t'|,|u'|\gg 1$, at fixed scattering angle, the fermion-loop induced amplitudes go to constants 
\be
\mathcal M^f_{++++}=1-\frac{(t'-u')}{s'}\left[\ell(t')-\ell(u')\right]
			+\frac{t'^2+u'^2}{2s'^2}\left([\ell(t')-\ell(u')]^2+\pi^2\right)+\dots\nn
\ee
\be
\mathcal M^f_{+++-}=\mathcal M^f_{++--}=-1+\dots
\ee
with the remaining amplitudes given by Eq.~(\ref{relation}) and we have defined the shorthand $\ell(z)\equiv\log(-1/4z)=-\log(-4z-i\epsilon)$.
The vector-loop induced amplitudes on the other hand show a logarithmic divergence  \cite{Jikia:1993tc,Gounaris:1999gh}.
\bea
\mathcal M^v_{++++}&=&-\frac{3}{2}+\frac{3}{2}\frac{(t'-u')}{s'}[\ell(t')-\ell(u')]
  -2\left(1-\frac{3}{4}\frac{t'u'}{s'^2}\right)\left([\ell(t')-\ell(u')]^2+\pi^2\right)\nn\\
 && -2s'^2\left(\frac{1}{s't'}\ell(s')\ell(t') +\frac{1}{s'u'}\ell(s')\ell(u') + \frac{1}{t'u'}\ell(t')\ell(u')\right)
\eea
\be
\mathcal M^v_{+++-}=
\mathcal M^v_{++--}=\frac{3}{2}+\dots
\ee
This has the important consequence that at LHC energies, the $W$-loop is  dominating the loops of all the SM fermions (including the top).  

\subsection{Forward and Backward limits}

For fermions, the forward limit ($|t'|\ll s'$) has also been computed in Ref.~\cite{Costantini:1971cj}:
\bea
\mathcal M^f_{++++}&=&1+\left(2-\frac{1}{s'}\right)B(s')+\left(-4+\frac{1}{s'}\right)B(-s')
+\left(\frac{1}{s'}-\frac{1}{2 s'^2}\right)T(s')\nn\\
	&&+\left(2-\frac{1}{s'}-\frac{1}{2s'^2}\right)T'(-s')\,,  \nn\\
\mathcal M^f_{++--}&=&-1-\frac{1}{s'}B(s')+\frac{1}{s'}B(-s')-\frac{1}{2s'^2}T(s')-\frac{1}{2s'^2}T(-s')\,,\label{fwdf}
\eea
as well as $\mathcal M^f_{+-+-}(s')=M^f_{++++}(-s')$ and $M^f_{+--+}=M^f_{+++-}=0$. Similarly, in the spin-1 case, we obtain
\bea
\mathcal M^v_{++++}&=&-\frac{3}{2}+\frac{8}{s'}\left(s'-\frac{1}{4}\right)
\left(s'-\frac{3}{4}\right)\left(B(s')-B(-s')+\frac{1}{2s'}T(s')+\frac{1}{2s'}T(-s')\right)\nn\\
	&&+3B(-s')
+\left(\frac{3}{s'}-8\right)T(-s')\,,  %\nn\\
%\mathcal M^v_{++--}&=&-1-\frac{1}{s'}B(s')+\frac{1}{s'}B(-s')-\frac{1}{2s'^2}T(s')-\frac{1}{2s'^2}T(-s')\,.
\label{fwdV}
\eea
while $M^v_{++--}=-\frac{3}{2}M^f_{++--}$, $\mathcal M^v_{+-+-}(s')=M^v_{++++}(-s')$, and $M^v_{+--+}=M^v_{+++-}=0$.
The backward limit ($|u'|\ll s'$) is obtained by the interchange $M_{+-+-}\leftrightarrow M_{+--+}$.
Notice that the forward/backward and high energy limits do not commute.

\section{Expressions for the amplitudes in EFT}
\label{app:C}

Starting from the effective Lagrangian, 
\be
\mathcal L_{\rm eff}=\zeta_1\ (F_{\mu\nu}F^{\mu\nu})^2+\zeta_2\ F_{\mu\nu}F^{\nu\rho}F_{\rho\lambda}F^{\lambda\mu} 
\ee
one can compute
\bea
 \alpha_{em}^2 \mathcal M_{++++}&=&-\frac{1}{4}\,(4\zeta_1+3\zeta_2)\,s^2\nn\\
 \alpha_{em}^2 \mathcal M_{++--}&=&-\frac{1}{4}\,(4\zeta_1+\zeta_2)\,(s^2+t^2+u^2)\nn\\
 \alpha_{em}^2 \mathcal M_{+++-}&=&\mathcal O(s'^3)\,.
\label{amplitudesEFTlimit}
\eea
In case the $\zeta_i$ arise in the low-energy limit of fermion and gauge loops Eq.~(\ref{EH}) these expressions precisely reproduce the low-energy limit of the amplitudes given in Eqs.~(\ref{LEf}) and (\ref{LEv}).

\bibliographystyle{JHEP} 

\bibliography{qgc_JHEP}

\end{document}